\documentclass[journal=ancac3,manuscript=article,layout=twocolumn]{achemso}
\usepackage{chemformula} % Formula subscripts using \ch{}
\usepackage[T1]{fontenc} % Use modern font encodings
\usepackage{graphicx}% Include figure files
\usepackage{dcolumn}% Align table columns on decimal point
\usepackage{bm}% bold math
\usepackage{placeins} % Allows use of FloatBarrier to keep figures in correct sections
\usepackage{amsmath} % For lots of math things
\usepackage{amssymb} % For some additional math symbols
\usepackage{csquotes} % For block (and other in-line) quotations - to handle repeated methods in experimental section
\usepackage{url}  % for url hyperlinks
\usepackage{braket} % For Dirac notation

\author{Brandon J. Furey}
\affiliation[University of Texas at Austin]
{Department of Physics, University of Texas at Austin, 2515 Speedway, C1600, Austin, TX USA 78712}
\email{furey@utexas.edu}
\author{Ben Stacy}
\affiliation[University of Texas at Austin]
{McKetta Department of Chemical Engineering, University of Texas at Austin, 200 E. Dean Keeton St., C0400, Austin, TX USA 78712}
\author{Tushti Shah}
\affiliation[University of Texas at Austin]
{McKetta Department of Chemical Engineering, University of Texas at Austin, 200 E. Dean Keeton St., C0400, Austin, TX USA 78712}
\author{Rodrigo M. Barba-Barba}
\affiliation[Centro de Investigaciones en \'{O}ptica]
{Centro de Investigaciones en \'{O}ptica, A.C., Loma del Bosque 115, Colonia Lomas del Campestre, Le\'{o}n, Gto., M\'{e}xico 37150}
\author{Ramon Carriles}
\affiliation[Centro de Investigaciones en \'{O}ptica]
{Centro de Investigaciones en \'{O}ptica, A.C., Loma del Bosque 115, Colonia Lomas del Campestre, Le\'{o}n, Gto., M\'{e}xico 37150}
\email{ramon@cio.mx}
\author{Alan Bernal}
\affiliation[Centro de Investigaciones en \'{O}ptica]
{Centro de Investigaciones en \'{O}ptica, A.C., Loma del Bosque 115, Colonia Lomas del Campestre, Le\'{o}n, Gto., M\'{e}xico 37150}
\author{Bernardo S. Mendoza}
\affiliation[Centro de Investigaciones en \'{O}ptica]
{Centro de Investigaciones en \'{O}ptica, A.C., Loma del Bosque 115, Colonia Lomas del Campestre, Le\'{o}n, Gto., M\'{e}xico 37150}
\author{Brian A. Korgel}
\affiliation[University of Texas at Austin]
{McKetta Department of Chemical Engineering, University of Texas at Austin, 200 E. Dean Keeton St., C0400, Austin, TX USA 78712}
\alsoaffiliation[Texas Materials Institute]{Texas Materials Institute, University of Texas at Austin, 204 E. Dean Keeton St., C2201, Austin, TX USA 78712}
\email{korgel@che.utexas.edu}
\author{Michael C. Downer}
\affiliation[University of Texas at Austin]
{Department of Physics, University of Texas at Austin, 2515 Speedway, C1600, Austin, TX USA 78712}
\email{downer@physics.utexas.edu}

\title[Two-Photon Excitation of Silicon Quantum Dots]
  {Two-Photon Excitation Spectroscopy of Silicon Quantum Dots and Ramifications for Bio-Imaging}


\begin{document}

%%%%%%%%%%%%%%%%%%%%%%%%%%%%%%%%%%%%%%%%%%%%%%%%%%%%%%%%
%% The "tocentry" environment can be used to create an entry for the graphical table of contents. It is given here as some journals require that it is printed as part of the abstract page. It will be automatically moved as appropriate.
%%%%%%%%%%%%%%%%%%%%%%%%%%%%%%%%%%%%%%%%%%%%%%%%%%%%%%%%

%\begin{tocentry}
%
%Some journals require a graphical entry for the Table of Contents.
%This should be laid out ``print ready'' so that the sizing of the
%text is correct.
%
%Inside the \texttt{tocentry} environment, the font used is Helvetica
%8\,pt, as required by \emph{Journal of the American Chemical
%Society}.
%
%The surrounding frame is 9\,cm by 3.5\,cm, which is the maximum
%permitted for  \emph{Journal of the American Chemical Society}
%graphical table of content entries. The box will not resize if the
%content is too big: instead it will overflow the edge of the box.
%
%This box and the associated title will always be printed on a
%separate page at the end of the document.
%
%\end{tocentry}

%%%%%%%%%%%%%%%%%%%%%%%%%%%%%%%%%%%%%%%%%%%%%%%%%%%%%%%%
%% The abstract environment will automatically gobble the contents if an abstract is not used by the target journal.
%%%%%%%%%%%%%%%%%%%%%%%%%%%%%%%%%%%%%%%%%%%%%%%%%%%%%%%%

\begin{abstract}
Two-photon excitation in the near-infrared (NIR) of colloidal nanocrystalline silicon quantum dots (nc-SiQDs) with photoluminescence also in the NIR has the potential to open up new opportunities in the field of deep biological imaging. Spectra of the degenerate two-photon absorption (2PA) cross section of colloidal nc-SiQDs are measured using two-photon excitation over a spectral range $1.46 < \hbar \omega < 1.91$ eV (wavelength $850 > \lambda > 650$ nm) above the two-photon band gap $E_g^{(QD)}/2$, and at a representative photon energy $\hbar \omega = 0.99$ eV ($\lambda = 1250$ nm) below this gap. Two-photon excited photoluminescence (2PE-PL) spectra of nc-SiQDs with diameters $d = 1.8 \pm 0.2$ and $d = 2.3 \pm 0.3$ nm, each passivated with 1-dodecene and dispersed in toluene, are calibrated in strength against 2PE-PL from a known concentration of Rhodamine B dye in methanol. The 2PA cross section is observed to be smaller for the smaller diameter nanocrystals and the onset of 2PA is observed to be blueshifted from the two-photon indirect band gap of bulk Si, as expected for quantum confinement of excitons. The efficiencies of nc-SiQDs for bio-imaging using 2PE-PL are simulated in various biological tissues and compared to other quantum dots and molecular fluorophores and found to be superior at greater depths.
\end{abstract}

%%%%%%%%%%%%%%%%%%%%%%%%%%%%%%%%%%%%%%%%%%%%%%%%%%%%%%%%
%% Start the main part of the manuscript here.
%%%%%%%%%%%%%%%%%%%%%%%%%%%%%%%%%%%%%%%%%%%%%%%%%%%%%%%%

\section{Introduction}

Numerous photonic applications for nanocrystalline semiconductor quantum dots have emerged in recent years, including spin qubits in photonic networks,\cite{kimdots,press,yamamoto,greve} quantum dot light-emitting diodes (LEDs),\cite{choi,yang,dai,pimputkar} \textit{in vitro} and \textit{in vivo} biological imaging,\cite{furey1,kim,chandra,kharin,sakiyama,mcvey,liang} and cancer therapy.\cite{kharin,tamarov,leencsi} Some of these applications use two-photon absorption (2PA) directly to excite photoluminescence (PL), taking advantage of the availability of ultrashort, high-peak-intensity laser sources and/or high sample transparency at the excitation wavelength. In other applications, 2PA is an undesired performance inhibitor that must be understood and managed. Prior measurements of 2PA in nc-SiQDs are either single wavelength measurements or do not characterize the sample sufficiently to extract a 2PA cross section, but rather only the 2PA dispersion.\cite{furey1,he,torrestorres,gui,falconieri,prakash,spano,trojanek} Here we measure two-photon excited photoluminescence (2PE-PL) spectra of colloidal nanocrystalline silicon quantum dots (nc-SiQDs)\cite{furey1,hessel2,yu} of diameters $d = 1.8 \pm 0.2$ and $d = 2.3 \pm 0.3$ nm over an excitation photon energy (wavelength) range $1.46 < \hbar \omega < 1.91$ eV ($650 < \lambda < 850$ nm) above the two-photon band gap $E_g^{(QD)}/2$ of the quantum dots, and at a representative photon energy (wavelength) $\hbar \omega = 0.99$ eV ($\lambda = 1250$ nm) below this gap. We extract 2PA cross sections over this range from the results. Nanocrystalline SiQDs luminesce efficiently in response to excitation in this range, a feature that bio-imaging applications such as two-photon excitation microscopy exploit.\cite{furey1,kim,chandra,kharin,tolstik,ravotto,tu} In these applications, nc-SiQDs offer nontoxicity,\cite{park,nontoxic} aqueous solution dispersibility,\cite{furey1,hessel2,yu} and size-dependent emission spectra\cite{brus,mastronardi} as advantages over photoluminescent dyes, while 2PE offers the advantage over one-photon excitation (1PE) of high tissue transparency and penetration depth at the excitation wavelength. Results presented here will aid in choosing excitation wavlength and nanocrystal size for bio-imaging applications, and will provide data for comparison with first-principles computations of 2PA in nc-SiQDs.

The electronic structure of nc-SiQDs consists of discrete energy levels as in molecules, rather than continuous bands as in bulk crystals.\cite{ramos1} The energy levels of nc-SiQDs originate primarily from quantum-confined excitons, but surface, interface,  and defect states can also contribute.\cite{ramos2} Exciton recombination dominates red and near infrared PL, while surface, interface, and defect states are believed to contribute to green and blue PL.\cite{ramos2} The latter, however, are not observed from the surface-passivated nc-SiQDs studied in this work. We will ascribe PL across the band gap of the quantum dots generically to a LUMO-HOMO transition, where LUMO (HOMO) refer to a lowest unoccupied (highest occupied) molecular orbital prior to excitation. For $d \gtrsim 2$ nm  diameter, the HOMO-LUMO gap and PL quantum yield both increase with decreasing $d$. This tunability of the PL spectrum is attractive for many applications, although it is limited in nc-SiQDs to $\hbar \omega_{PL} < 2.1$ eV.\cite{wen}

Selection rules for excitation by one-photon absorption (1PA) and 2PA can differ. In centrosymmetric molecules they are mutually exclusive, since 1PA electric dipole transitions are parity-forbidden while their 2PA counterparts are parity-allowed. In this case, 1PA and 2PA excite complementary states.\cite{diener} However, at room temperature the PL spectra and PL quantum yields of nc-SiQDs are nearly the same for both modes of excitation.\cite{diener} Therefore, monitoring the spectrally-integrated 2PE-PL spectra as a function of incident light intensity can indirectly be used to measure the 2PA cross section.\cite{rumi}

\section{Results and Discussion}

\subsection{Characterization of 1-Dodecene-Passivated Colloidal nc-SiQDs}

\begin{figure*}
  \includegraphics[width=\textwidth]{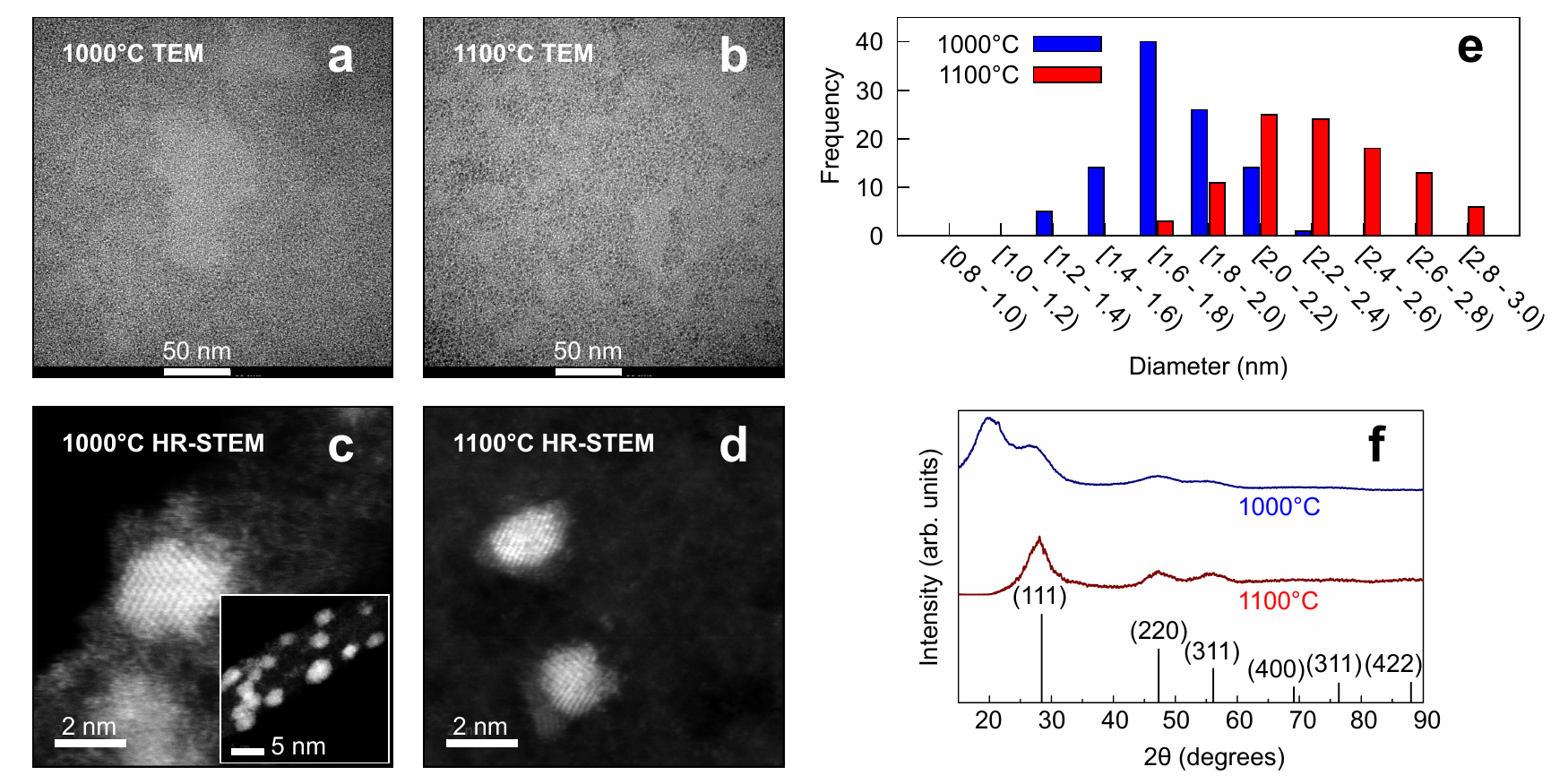}
  \caption{TEM images of nc-SiQDs annealed at $1000^{\circ}$C (a) and $1100^{\circ}$C (b). Aberration-corrected high-resolution dark field STEM images of nc-SiQDs annealed at $1000^{\circ}$C (c), with inset zoomed out showing the quantum dots attached to the graphene grid, and $1100^{\circ}$C (d). Histogram of size distributions from TEM images of each nc-SiQD sample are shown in (e).  XRD of nc-SiQD samples indicate crystallinity due to the presence of (111), (220), and (311) diffraction peaks in $m3m$ crystalline Si as shown in (f).}
  \label{fig:siqdtemchar}
\end{figure*}

\begin{figure}
  \includegraphics[scale=1.0]{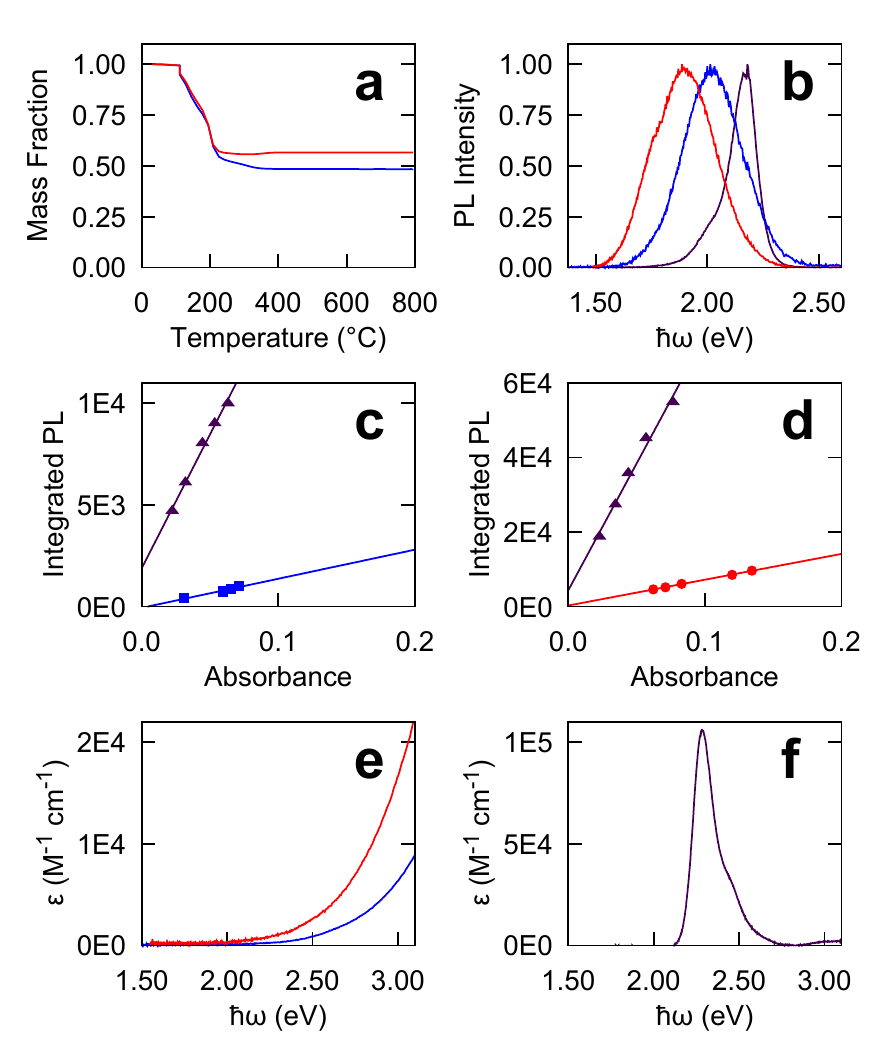}
  \caption{Characterization of nc-SiQD sample $d = 1.8$ nm (blue), nc-SiQD sample $d = 2.3$ nm (red), and the reference standard RhB (violet): (a) Mass fraction as a function of temperature of nc-SiQD samples determined by TGA, (b) normalized one-photon excited PL excited at $\hbar \omega = 3.26$ eV, (c) quantum yield of $d = 1.8$ nm nc-SiQDs determined by integrated one-photon excited PL (in arb. units) as a function of absorbance relative to RhB, (d) quantum yield of $d = 2.3$ nm nc-SiQDs determined by integrated one-photon excited PL (in arb. units) as a function of absorbance relative to RhB, (e) molar absorptivity of nc-SiQD samples, (f) molar absorptivity of RhB from literature.\cite{du}}
  \label{fig:siqdotherchar}
\end{figure}

The nc-SiQDs studied in this work were synthesized by thermal decomposition of hydrogen silsesquioxane at annealing temperatures of $1000^{\circ}$C and $1100^{\circ}$C which yielded nc-SiQDs with size distributions of $d =  1.8 \pm 0.2$ nm and $2.3 \pm 0.3$ nm, respectively, as determined by transmission electron microscopy (TEM) [see Figure \ref{fig:siqdtemchar}(a)-(b), and (e)]. The nc-SiQDs were passivated with 1-dodecene and dispersed in toluene. The Si quantum dots are confirmed to be crystalline by the presence of (111), (220), and (311) diffraction peaks for each of these crystal planes in $m3m$ crystalline Si by X-ray diffraction (XRD) [see Figure \ref{fig:siqdtemchar}(f)]. The broad peak near $20^{\circ}$ seen in the nc-SiQD sample annealed at $1000^{\circ}$C but not in the sample annealed at $1100^{\circ}$C is likely due to the 1-dodecene ligand. This broad peak is expected to be more visible in the $1000^{\circ}$C nc-SiQD sample due to the lower mass fraction of Si core to 1-dodecene. Thermal gravimetric analysis (TGA) was performed in order to determine the relative mass of the Si core to 1-dodecene. This is necessary to relate the mass concentration of a sample to a number density such that a 2PA cross section per quantum dot can be calculated. The mass fraction \textit{vs.} temperature curves are shown in Figure \ref{fig:siqdotherchar}(a). One-photon excited PL (1PE-PL) of nc-SiQDs and Rhodamine B (RhB), excited at $\hbar \omega = 3.10$ eV, are shown in Figure \ref{fig:siqdotherchar}(b). The quantum yields ($\phi_{PL}$) of the nc-SiQDs were determined by comparing the integrated 1PE-PL emission spectra to that of a known reference standard, RhB in anhydrous ethanol. Integrated PL is plotted against absorbance for various concentrations of both nc-SiQDs and RhB, shown in Figure \ref{fig:siqdotherchar}(c)-(d). The ratio of the gradient of the trendline of luminescence intensity \textit{vs.} absorbance of nc-SiQDs to that of RhB is proportional to the quantum yield. The calculated quantum yields for the nc-SiQDs in this study are $\phi_{PL} = 0.064$ for $d = 1.8 \pm 0.2$ nm and $\phi_{PL} = 0.060$ for $d=2.3 \pm 0.3$ nm. The molar absorptivities of the nc-SiQD samples were calculated and are shown in Figure \ref{fig:siqdotherchar}(e) and the molar absorptivity of RhB from Du \textit{et al.}\cite{du} is shown in Figure \ref{fig:siqdotherchar}(f).

\subsection{2PE-PL Measurements}

\begin{figure}
  \includegraphics[scale=1.0]{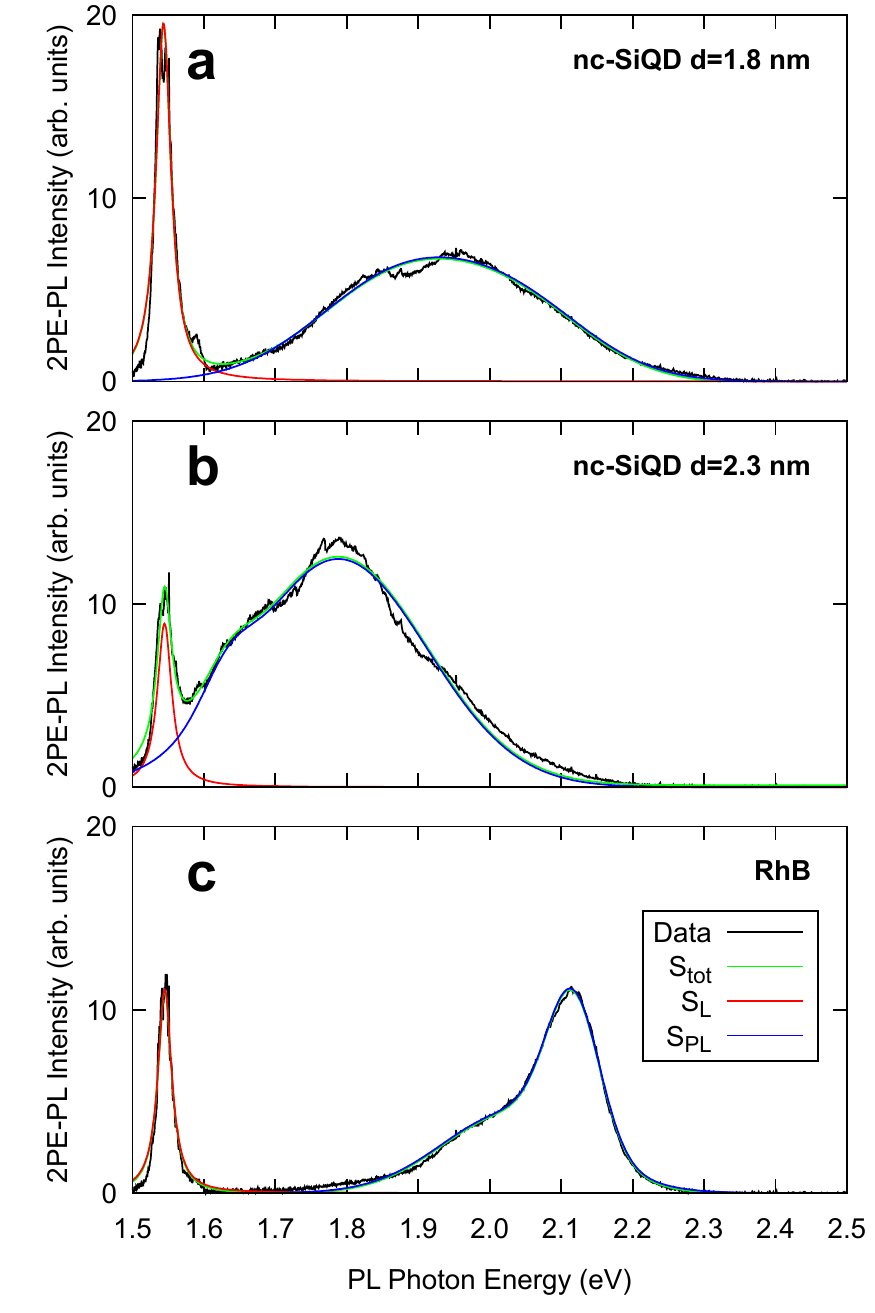}
  \caption{2PE-PL spectra observed at excitation $\hbar \omega = 1.55$ eV (black) with total fit (green) accounting for the laser lineshape (red) and 2PE-PL spectra (blue). a) nc-SiQD sample $d=1.8 \pm 0.2$ nm with incident pulse energy $\varepsilon_P = 3.735 \pm 0.008\ \mu$J, b) nc-SiQD sample $d=2.3 \pm 0.3$ nm with $\varepsilon_P = 3.804 \pm 0.008\ \mu$J, c) RhB dye with $\varepsilon_P = 3.712 \pm 0.008\ \mu$J.}
  \label{fig:2pepspec}
\end{figure}

We two-photon excited PL using laser pulses from an optical parametric amplifier (OPA). The output power of the OPA is sufficient to observe 2PE-PL except in the range $0.99 < \hbar \omega < 1.46$ eV, where operation switches from second-order frequency mixing of the signal to second harmonic generation (SHG) from the idler. The PL spectra of the samples were recorded as a function of incident pulse energy. An example of 2PE-PL spectra for the nc-SiQD samples and RhB dye excited at $\hbar \omega = 1.55$ eV is shown in Figure \ref{fig:2pepspec}. The 2PE-PL spectral shape did not differ from 1PE-PL (see the Supporting Information). The observed spectra $S_{tot}(E) = S_L(E) + S_{PL}(E)$ includes scattered laser light in addition to the PL emitted from the sample, and since the laser lineshape sometimes overlaps with the PL spectra, this has to be removed. The PL spectra $S_{PL}(E)$ is proportional to the PL photon number within a photon energy bin of width $\mathrm{d}E$ centered at $E$. It can be empirically modeled as a sum of two Gaussian peaks and the laser lineshape $S_L(E)$ is modeled as a Lorentzian. The total detected spectra is then described by the sum of these spectral lineshapes. 

The total number of detectable emitted PL photons is given by
\begin{equation}
\label{eqnumberphotonsintegral}
N_{PL} = \int_{E_-}^{E_+} S_{PL}(E)\ \mathrm{d}E,
\end{equation}
where $E_- = 1.24$ eV and $E_+ = 6.20$ eV are the detector limits for the spectrometer. The lineshape fitting parameters for the samples studied in this work are tabulated in the Supporting Information.

The number $N_{PL}$ of detected PL photons from a single excitation pulse is directly related to the excited state population $N_e$ after the pulse has propagated through the sample, $N_{PL} = f_{det} \phi_{PL} N_{e}$, where $f_{det}$ characterizes the collection efficiency of the detector and $\phi_{PL}$ is the PL quantum yield. The excited state population is driven by 2PA for which the attenuation of the incident beam with propagation distance $z$ is quadratic in its incident intensity $I(z,r,t)$,
\begin{equation}
\label{eq2padiffeq}
\frac{\mathrm{d}I(z,r,t)}{\mathrm{d}z} = - \beta \big[ I(z,r,t)\big]^2,
\end{equation}
where $\beta = \sigma n_g$ is the 2PA coefficient, $\sigma$ is the 2PA cross section, and $n_g$ is the ground state population density. Here we neglect free-carrier and linear absorption. $\phi_{PL}$ for the samples can be assumed to be the same as measured using one-photon excitation since the PL emission spectra of the two excitation channels do not appreciably differ.\cite{diener,rumi} The value of $f_{det}$ can then be determined from fitting the intensity dependence of the PL from a reference standard for which $\phi_{PL}$, $\sigma$, and $n_0$ are known. Thus $\sigma$ for the samples can be determined relative to the reference standard.

We can model the excitation of nc-SiQDs by 2PA as an effective two-level molecular system as long as the populations of higher energy excited states rapidly transition to the LUMO. The time rates-of-change of the excited ($n_e$) and ground state population densities are described by
\begin{align}
\label{eqnediffeq}
\frac{\mathrm{d}n_e(z,r,t)}{\mathrm{d}t} &= \frac{\beta \big[I(z,r,t)\big]^2}{2 \hbar \omega} - \Gamma n_e(z,r,t)\\
\label{eqn0diffeq}
\frac{\mathrm{d}n_g(z,r,t)}{\mathrm{d}t} &= - \frac{\mathrm{d}n_e(z,r,t)}{\mathrm{d}t},
\end{align}
where $\hbar \omega$ is the photon energy of the excitation pulse and $\Gamma$ is the recombination rate. The initial conditions are $n_e(z,r,-\infty) = 0$ and $n_g(z,r,-\infty) = n_0$, where $n_0$ is the nc-SiQD number density and $t = 0$ corresponds to the arrival of the driving pulse peak. The recombination times of the samples in this work are on the order of ns -- $\mu$s, which is much longer than the timescale of a laser pulse of $\tau_g \sim 100$ fs but much shorter than the laser repetition period of $1/f_{rep} = 1$ ms, \textit{i.e.,} $f_{rep} \ll \Gamma \ll 1 / \tau_g$. Thus recombination can be neglected on the timescale of a pulse and the sample can be considered to be fully relaxed before the arrival of a subsequent pulse. 

We assume a Gaussian spatiotemporal laser pulse profile with peak on-axis intensity $I_0$ and a sample thickness shorter than the Rayleigh range. The solution to Equation \ref{eqnediffeq} for this intensity profile, neglecting depletion of the ground state and pump beam, can then be integrated over volume to find the excited state population. This can then be related to the number of detected PL photons, which are emitted as excitons radiatively recombine across the HOMO-LUMO gap. This ``lowest-order'' solution is quadratic with incident intensity.

\begin{figure}
  \includegraphics[scale=1.0]{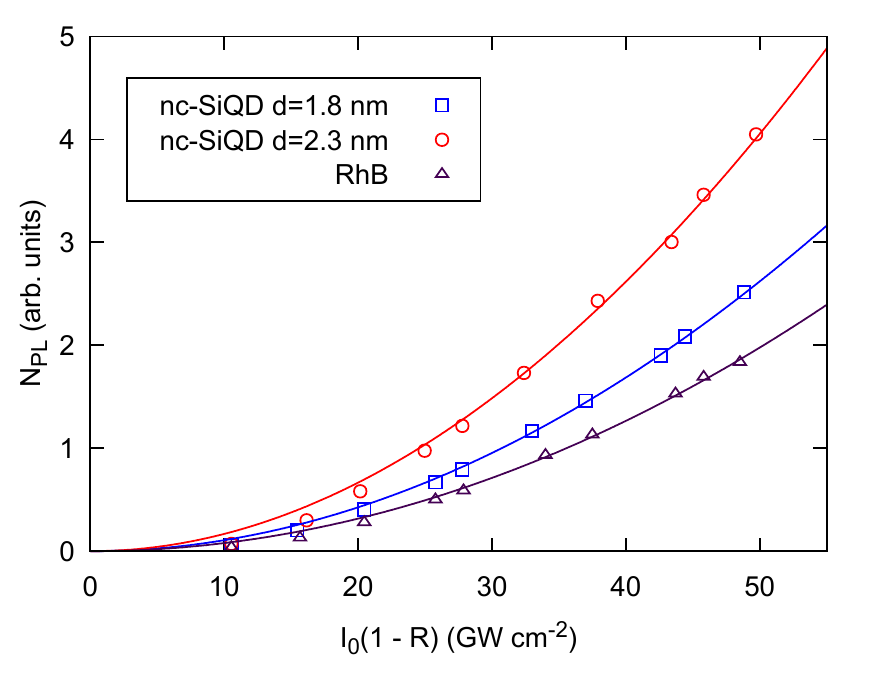}
  \caption{$N_{PL}$ \textit{vs.} $I_0(1 - R)$ at $\hbar \omega = 1.55$ eV for nc-SiQD sample $d = 1.8$ nm (blue squares), nc-SiQD sample $d = 2.3$ nm (red circles), and reference standard RhB (violet triangles), and corresponding fits to Equation \ref{eqnplbetapumpdepletion} with $N = 2$ (blue curve), $N = 2$ (red curve), and $N = 1$ (violet curve).}
  \label{fig:tpepintphi0}
\end{figure}

We can calculate figures of merit for the validity of this model by estimating the fraction of absorbed photons on the beam axis $\Delta_{ph} \approx \sigma n_0 z I_0 / \sqrt{2}$ and the fraction of excited molecules on the beam axis $\Delta_{m} \approx \sqrt{\pi} \sigma I_0^2 \tau_g / (4 \hbar \omega)$,\cite{rumi} where $w$ is the beam radius and $\tau_g = \tau_{FWHM}/\sqrt{2 \mathrm{ln}2}$ is the Gaussian pulse duration parameter expressed in terms of the FWHM pulse duration. In this work, we find that $\Delta_m < 3\%$ always, and thus neglecting depletion of the ground state is a valid assumption. However, $\Delta_{ph} < 3\%$ sometimes, but not always. In cases where pump depletion is significant, we can expand the solution about $\beta z I_0 \approx 0$ to higher orders giving
\begin{multline}
\label{eqnplbetapumpdepletion}
N_{PL} = - \frac{\pi^{3/2} f_{det} \phi_{PL} w^2 \tau_g I_0}{2^{5/2} \hbar \omega}\\
 \lim_{N \rightarrow \infty} \sum_{j=1}^{N} \frac{(-\sigma n_0 z I_0)^j}{(j+1)^{3/2}},
\end{multline}
where the higher order terms account for pump depletion. The data is fit to Equation \ref{eqnplbetapumpdepletion} with $N = 1$ and again with $N = 2$, and if the variation in the fit parameter $\sigma$ between each fit is below a threshold of 1\%, then the result for the $N = 1$ case is reported; if the variation is larger, then the fit is repeated, comparing successively higher-order terms until the fit parameter converges. An example of this analysis is shown in Figure \ref{fig:tpepintphi0} at $\hbar \omega = 1.55$ eV. The fit parameters are tabulated in the Supporting Information.

\subsection{2PA Cross Section Spectra}

\begin{figure*}
  \includegraphics[width=\textwidth]{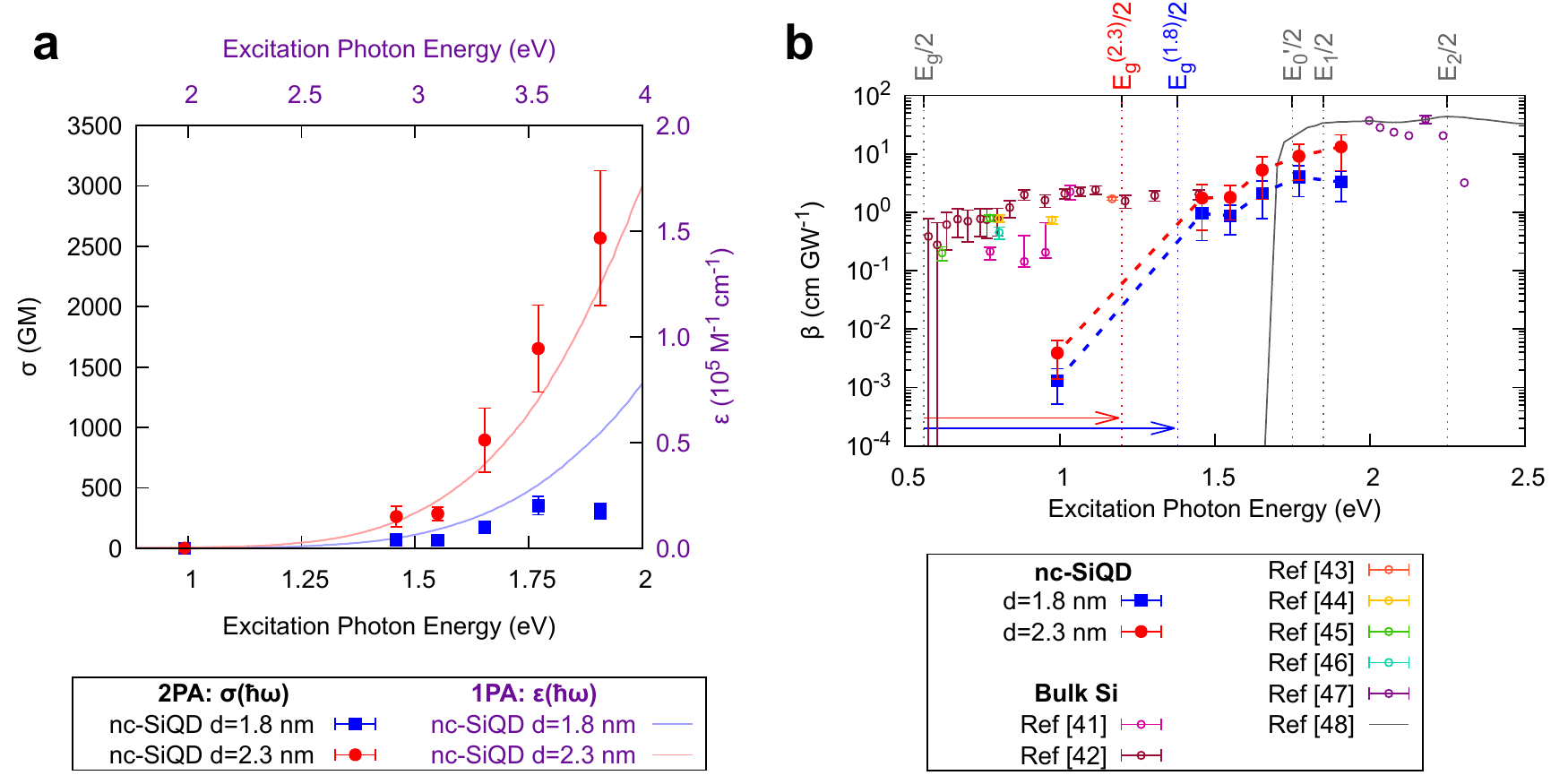}
  \caption{(a) 2PA cross section $\sigma(\hbar \omega)$ spectra in units of GM (1 G\"{o}ppert-Mayer = 10$^{-50}$ cm$^4$ s photon$^{-1}$) for nc-SiQDs (red and blue points), measured relative to RhB, then converted to absolute values using RhB cross sections from Makarov \textit{et al.}\cite{makarov}. The molar absorptivity $\varepsilon(\hbar \omega)$ spectra of the nc-SiQDs is plotted on the right axis against the top axis (light red and blue curves on violet axes). (b) Degenerate 2PA coefficient $\beta(\hbar \omega)$ spectra in units of cm GW$^{-1}$ for nc-SiQDs with volume fraction $V_{frac} = 1$ (red and blue solid points). Dashed red and blue lines are guides to the eye. Experimental results for bulk Si in literature are shown in colored open circles.\cite{furey2, bristow,reintjes,dinu,euser,tsang,reitze} Transitions in bulk Si between $E_g/2 < \hbar \omega < E_0'/2$ are indirect. The blueshifts of the 2PA band edges of the nc-SiQD samples from that of bulk Si are indicated with horizontal arrows between $E_g/2$ for bulk Si and the empirically-derived HOMO-LUMO gaps for nc-SiQDs ($E_g^{(1.8, 2.3)}$). \textit{Ab initio} calculations for bulk Si in literature for direct transitions above $E_0'$ are also shown (gray line)\cite{murayama} along with the $E_1/2$ and $E_2/2$ critical points.}
  \label{fig:sigmaspec}
\end{figure*}

The values of the 2PA cross sections $\sigma$ of the nc-SiQD samples relative to the RhB reference standard were measured over the range $0.99 < \hbar \omega < 1.91$ eV. The resulting spectra, based on RhB 2PA cross sections from Makarov \textit{et al.},\cite{makarov} are shown in Figure \ref{fig:sigmaspec}(a). The values and their standard errors are tabulated in the Supporting Information. The error bars are propagated from the statistical uncertainties in the fit parameters, the reported uncertainties in the 2PA cross section of the reference RhB,\cite{makarov} and the statistical uncertainties in the measured number densities (refer to the Supporting Information). The 2PA cross section of nc-SiQDs increases by almost a factor of 1000 between 0.99 eV and 1.46 eV, and continues to monotonically increase by about a factor of 10 between 1.5 and 1.9 eV. The 1PA molar absorptivity $\varepsilon(\hbar \omega )$ spectra are also shown in Figure \ref{fig:sigmaspec}(a), plotted on the right axis against the top axis for comparison of the dispersion to the 2PA spectra. The absorption in both channels increases monotonically with excitation photon energy, but differs in quantitative details.

The 2PA cross section can be related to an effective 2PA coefficient $\beta = \sigma n_0$ for a material with volume fraction $V_{frac} = n_0 V_{QD} = 1$, where $n_0$ is the quantum dot number density and $V_{QD} = (\pi/6) d^3$ is the volume of the core of a quantum dot of diameter $d$. This can then be used to compare how the 2PA spectra of nc-SiQDs compare to bulk Si [refer to Figure \ref{fig:sigmaspec}(b)]. Bulk Si has a two-photon indirect band gap of  $E_g /2 = 0.56$ eV and two-photon direct band gap of $E_0'/2 = 1.75$ eV. The 2PA cross section of nc-SiQDs increases by a factor of $\approx 10$ between 1.5 and 1.9 eV, before the two-photon direct band gap of bulk Si. The large increase in 2PA cross section of the nc-SiQDs between 0.99 eV and 1.46 eV is consistent with a blueshifting in the onset of absorption with decreasing nanocrystal size, which in the limit of bulk Si occurs at the two-photon indirect band gap. The HOMO-LUMO gap has been empirically observed to follow the trend $E_g \approx E_g^{\mathrm{bulk}} + 2.96/d$, for $E_g$ in eV, $d$ in nm, and where $E_g^{\mathrm{bulk}} = 1.12$ eV is the indirect band gap of bulk Si.\cite{ramos2} This trend holds for the onset of linear absorption in experimental measurements\cite{hessel2,ledoux,vanbuuren,furukawa,hessel} and \textit{ab initio} calculations.\cite{ramos2,oegut,reboredo} This relation gives a rough estimation for the onset of 2PA of $E_g^{(1.8)}/2 \approx 1.38$ eV for $d = 1.8$ nm nc-SiQDs and $E_g^{(2.3)}/2 \approx 1.20$ eV for $d = 2.3$ nm nc-SiQDs. These estimates for the HOMO-LUMO gap fall within the range $0.99 < \hbar \omega < 1.46$ eV where 2PA increased by a factor of almost 1000 and thus is consistent with our results.

The magnitude of the 2PA cross section at a given excitation energy is smaller for the smaller nanocrystals in this study, \textit{i.e.,} $\sigma^{(1.8)}(\omega) < \sigma^{(2.3)}(\omega)$, where the superscript indicates the diameter of the nc-SiQDs in nm [see Figure \ref{fig:sigmaspec}(a)]. This is consistent with the trend of linear molar absorptivity dependence on nc-SiQD size\cite{ramos2,hessel} as well as previous measurements of the 2PA cross section dependence on nc-SiQD size.\cite{furey1} Interestingly, the magnitude of the 2PA cross section may scale faster than nanocrystal volume, although the error bars for the two sizes of nanocrystals do overlap. This is evident in Figure \ref{fig:sigmaspec}(b) as the effective 2PA coefficient $\beta^{(2.3)}(\omega) > \beta^{(1.8)}(\omega)$.

\subsection{Comparing the Simulated Efficiency of 2PE-PL Biological Imaging with nc-SiQDs to Other Imaging Agents}

\begin{figure}
  \includegraphics[scale=0.8]{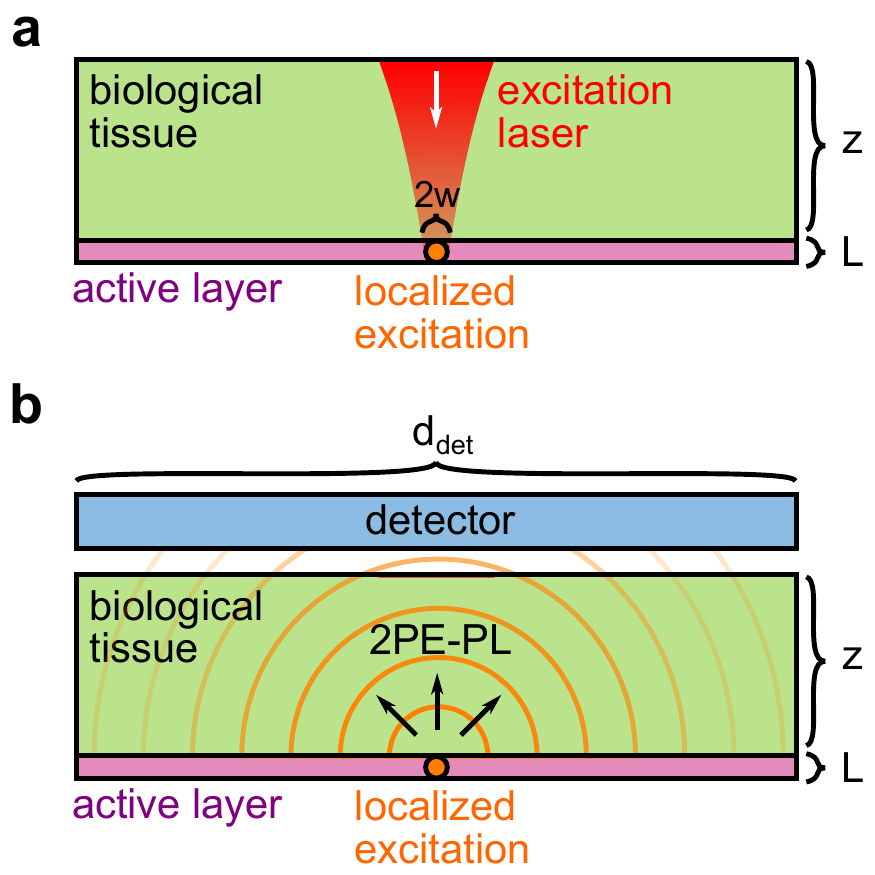}
  \caption{Model used in calculations of $q(z)$ of various imaging agents for biological imaging using 2PE-PL. An excitation laser beam transmits through the biological tissue (a) and two-photon excites a localized region near the focal volume in the active layer containing imaging agents. This localized excitation then emits isotropic 2PE-PL (b) which then transmits back to the surface of the biological tissue to be collected by a detector.}
  \label{fig:biomodel}
\end{figure}

The onset of 2PA in the near-infrared (NIR) bio-transparency window, the size-tunability of the PL spectra, and the non-toxicity of nc-SiQDs make them potentially attractive for deep bio-imaging applications alongside new organic dyes\cite{schnermann,strack,hemmer} and other quantum dots.\cite{pu,padilha,cai,borsella} We can simulate the efficiency of nc-SiQDs in generating an observable signal with 2PE-PL in biological tissue and compare to the expected signal from 2PE-PL using other quantum dots and molecular fluorophores to aid in the selection of the optimal imaging agent and excitation wavelength for a given application (see Figure \ref{fig:biomodel}). This simulation models a parameter $q(z)$ from which the maximum number of 2PE-PL photons which are returned to the surface of a biological sample can be calculated from the lowest-order term of Equation \ref{eqnplbetapumpdepletion} as a function of depth $z$ of the imaging agent in the tissue, $
\max_{E_{excite}} N_{PL}^{(surface)}(z) = (\pi/16) f_{det}(E_0) n_0 L w^2 \tau_g I_0^2 q(z)$, where $E_{excite}$ is the excitation photon energy, $f_{det}(E_0) = \phi_{det}(E_0) \Omega_{det}/(2\pi)$, $\phi_{det}(E_0)$ is the detector quantum efficiency, $\Omega_{det}$ is the detector collection solid angle, and $L$ is the thickness of the active layer where imaging agents are present. This model assumes that PL emission radiates from a point source at the focal volume and that $d_{det} \gg z \gg w,L$, where $d_{det}$ is the detector diameter. The detector quantum efficiency is assumed not to vary rapidly over the PL spectra, and we model the PL spectral irradiance as Gaussian peaks with a center position at $E_0$ and peak width $\Delta E_0$. 

We can express the parameter $q(z)$ in units of J$^{-2}$ cm$^4$ s as
\begin{multline}
\label{eqbioimagingcalc2}
q(z) = \max_{E_{excite}} \Bigg\{ \frac{\phi_{PL} \sigma(E_{excite}) e^{-2\alpha_{bio}(E_{excite})z}}{E_{excite} \Delta E_0 \Big[1 + \mathrm{erf}\big(\frac{E_0}{\Delta E_0}\big)\Big]}\\
 \int_{E_{IRCO}}^{\infty} \frac{e^{-(E - E_0)^2/\Delta E_0^2}}{E}\\
 \bigg( \int_0^{\pi/2} e^{-\alpha_{bio}(E) z/\cos{\theta}} \sin{\theta}\ \mathrm{d}\theta\bigg)\ \mathrm{d}E\Bigg\},
\end{multline}
where $\alpha_{bio}(E)$ is the effective linear attenuation coefficient of the biological tissue (accounting for losses due to absorption and scatter) and $\theta$ is the angle to the normal. The prefactor $\phi_{PL} \sigma(E_{excite})$ is the 2PE-PL efficiency at $E_{excite}$, $e^{-2\alpha_{bio}(E_{excite}) z}$ is the linear attenuation in the biological tissue of the square of the incident intensity, $E_{excite}$ appears in the denominator in the conversion of $\sigma$ from GM to SI units, the $\Delta E_0 \Big[1 + \mathrm{erf}\big(\frac{E_0}{\Delta E_0}\big)\Big]$ term in the denominator is a normalization factor for the integral immediately following, and $E_{IRCO}$ is a small but finite infrared cutoff energy. The first integral performed over $\mathrm{d}\theta$ represents the fraction of PL at a given photon energy $E$ which is emitted into the upper hemisphere and which transmits to the surface of the biological tissue of depth $z$. The next integral over $\mathrm{d}E$ is the fraction of emitted photons which are transmitted to the surface and accounts for the PL spectrum over which $\alpha_{bio}(E)$ may vary. 

\begin{figure*}
  \includegraphics[width=\textwidth]{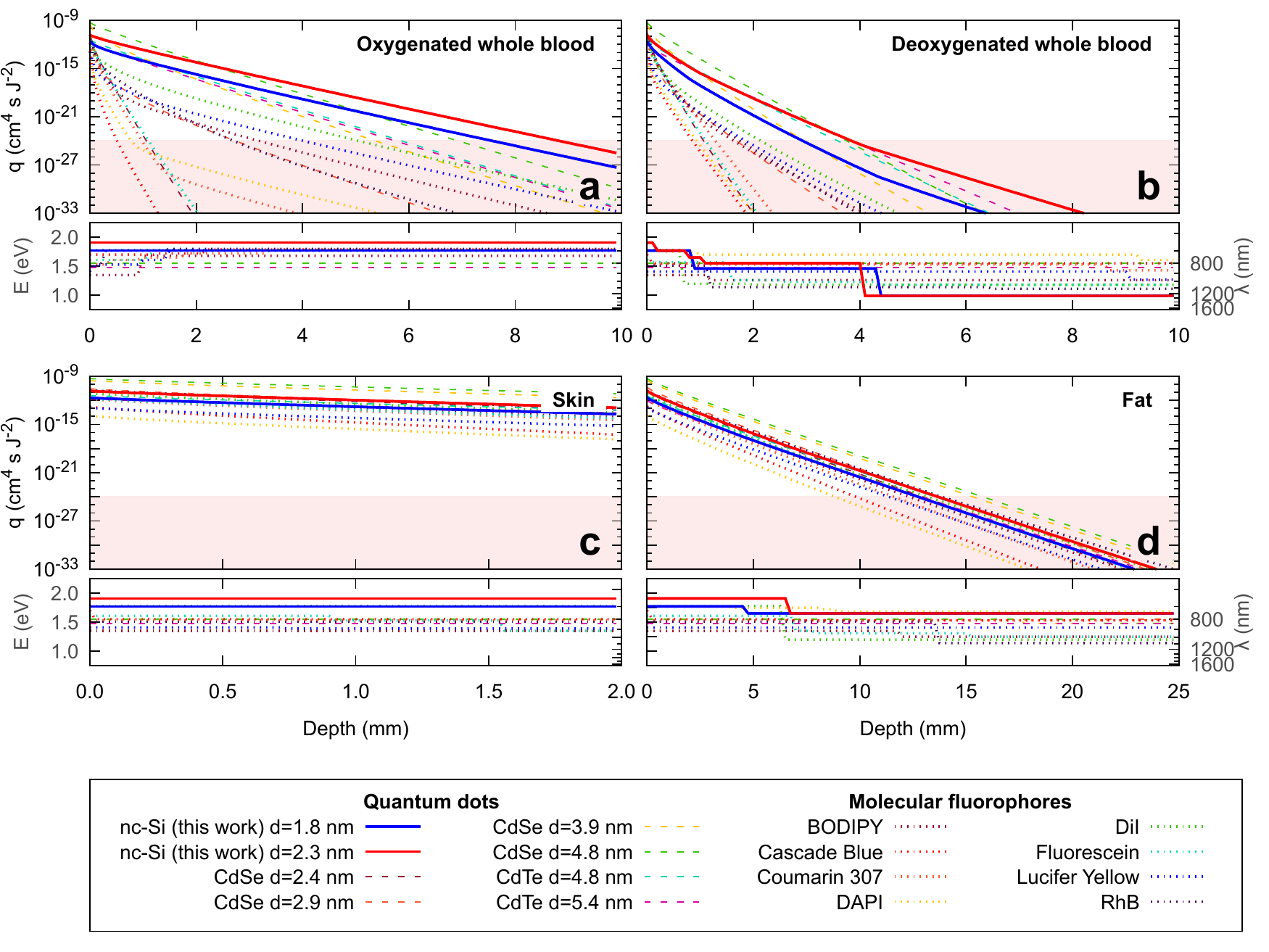}
  \caption{Simulated $q(z)$ excited by 2PA (upper subplot) and $E_{excite},\ \lambda_{excite}$ for $\max_{E_{excite}} N_{PL}^{surface}(z)$ (lower subplot) for various samples as a function of depth $z$ into biological tissue; (a) oxygenated whole blood, (b) deoxygenated whole blood, (c) skin, and (d) fat. The red shaded region indicates where the returned signal is below the approximate detection threshold for typical best-case scenarios (acquisition of a single photon over $\approx 1$ s with near unitary $f_{det}$, $f_{rep} = 1$ kHz, $n_0 \approx 10^{18}$ cm$^{-3}$, $L \approx 1$ mm, $w \approx 100\ \mu$m, $\tau_g \approx 100$ fs, and $I_0 \approx 100$ GW cm$^{-2}$).}
  \label{fig:bioimagingcalculations}
\end{figure*}

The results of these simulations for the nc-SiQDs in this work, CdSe and CdTe quantum dots, and a variety of molecular fluorophores are shown in Figure \ref{fig:bioimagingcalculations}. The nc-SiQDs show superior efficiency, or more precisely exhibit large $q(z)$ and the least reduction in signal with depth, $|\mathrm{d}q(z)/\mathrm{d}z|$, in oxygenated whole blood and deoxygenated whole blood than the other quantum dots and molecular fluorophores simulated. Compared to the molecular fluorophores simulated, nc-SiQDs can produce a detectable signal at an approximate threshold for detection $\approx 2 - 8 \times$ deeper. Even within skin and fat, nc-SiQDs offer high performance, especially for deep bio-imaging. This is primarily due to two factors related to the deeper penetration depth of longer wavelength light in biological tissue; the PL emission peak of nc-SiQDs is in the NIR, and the 2PA cross section is larger in the NIR. The results presented here open up new opportunities, particularly in the field of deep bio-imaging.

\section{Conclusions}

In conclusion, we characterized the 2PA cross section spectra of two different sizes of ligand-passivated and colloidally dispersed nc-SiQDs with diameters $d=1.8 \pm 0.2$ and $d=2.3 \pm 0.3$ nm using 2PE-PL as a proxy relative to the well-known reference standard of RhB over the range $0.99 < \hbar \omega < 1.91$ eV. We observed the 2PA cross section decreases with decreasing nanocrystal diameter. The HOMO-LUMO gap in the nc-SiQDs was blueshifted from the two-photon indirect band gap of bulk Si, as expected due to quantum confinement of excitons, matching the trend observed in linear absorption experiments\cite{hessel2,ledoux,vanbuuren,furukawa,hessel} and \textit{ab initio} calculations.\cite{ramos2,oegut,reboredo}. \textit{Ab initio} calculations of the rotationally-averaged imaginary part of the molecular third-order nonlinear optical susceptibility tensor of nc-SiQDs as a function of size could be compared to experimentally measured 2PA cross sections, and such calculations would be useful for understanding both the 2PA spectral structure and dependence on nanocrystal size. The efficiencies of nc-SiQDs for bio-imaging using 2PE-PL were simulated in various biological tissues and compared to other quantum dots and molecular fluorophores and found to be superior at greater depths due to their NIR PL and 2PA peaks, offering a potential advancement to the field of deep bio-imaging.

\FloatBarrier % This was added for the arXiv preprint v1 submission to force the bioimaging figure into the main body of the text
\section{Methods}

\subsection{Sample Preparation}

Hydrogen-passivated nanocrystals were synthesized by thermally decomposing hydrogen silsesquioxane, mechanically grinding the product into oxide-embedded nc-SiQDs, and then suspending the powder in an acid solution to etch away the oxide. This procedure yields hydride-terminated nc-SiQDs. Hessel \textit{et al.}\cite{hessel} and the Supporting Information describe the procedure in detail. The hydride-terminated nc-SiQD sample is then dispersed in 8 mL of 1-dodecene. The dispersion is heated at 190$^{\circ}$C for 20 h. After about 30 min of heating, the turbid brown dispersion turns to an optically clear orange dispersion, indicating passivation of the nc-SiQDs. The alkene passivated nc-SiQDs are washed three times by precipitation with ethanol as an antisolvent (15 mL) and redispersed in 1 mL of toluene for use in experiments. Refer to the Supporting Information for details of the alkene passivation procedure. The reference standard for 2PE-PL measurements in this work was RhB dye in methanol for which a fresh sample was prepared each day measurements were taken.

\subsection{Sample Characterization}

\subsubsection{Transmission Electron Microscopy}

Average nanocrystal diameters and size distributions were determined by imaging individual nanocrystals with a high-resolution TEM (JEOL Ltd. Model 2010F HR-TEM) operated at 200 kV. The particle size distributions were determined by calculating the sizes of 100 particles in each sample. See Figures \ref{fig:siqdtemchar}(a)-(b) and (e). Graphene-enhanced lacey carbon TEM grids were purchased from Electron Microscopy Sciences (Cat. no. GF1201). A dilute nanocrystal solution in chloroform was dropcast onto the grid and stored in a vacuum chamber overnight before imaging. Aberration-corrected scanning TEM (acSTEM) was performed using a JEOL NEOARM TEM with an 80 kV accelerating voltage and a point-to-point STEM resolution of 0.11 nm. These high resolution acSTEM images are shown in Figures \ref{fig:siqdtemchar}(c)-(d).

\subsubsection{X-Ray Diffraction}

XRD was performed on each of the nc-SiQD samples using a Rigaku R-Axis Spider diffractometer using Cu K$\alpha$ radiation ($\lambda = 0.15418$ nm) to ascertain the crystallinity of the quantum dots. The quantum dots were deposited on a glass slide, the solvent was evaporated, and then the powder was placed on a nylon loop. Two-dimensional diffraction data were collected for 10 min while rotating the sample stage at $10^{\circ}$ per minute. 2D diffraction data were radially integrated with 2DP software and are shown in Figure \ref{fig:siqdtemchar}(f).

\subsubsection{Thermal Gravimetric Analysis} 

TGA was performed using an automated ultra-micro balance (Mettler-Toledo International, Inc. Model TGA-1) in order to determine the relative mass of the Si core to 1-dodecene. Samples were heated at a rate of 20$^{\circ}$C/min from 40$^{\circ}$C to 800$^{\circ}$C. The sample was held at 100$^{\circ}$C for 30 minutes to evaporate residual solvents and at 800$^{\circ}$C for 30 min to ensure all ligand had evaporated. The relative mass of Si core to 1-dodecene in the passivated samples was determined from the remaining mass of Si and the total weight loss due to the removal of the ligands. The mass fraction \textit{vs.} temperature curves are shown in Figure \ref{fig:siqdotherchar}(a).

\subsubsection{Quantum Yield Calculations}

1PE-PL, shown in Figure \ref{fig:siqdotherchar}(b), and PL excitation (PLE) spectra in the ultraviolet--visible (UV--Vis) wavelength range were acquired on a fluorescence spectrophotometer (Varian, Inc. Model Cary Eclipse). UV--Vis absorbance spectroscopy was performed on a UV--Vis spectrophotometer (Varian, Inc. Model Cary 50 Bio UV--Vis). The quantum yields ($\phi_{PL}$) of the nc-SiQDs were determined by comparing the integrated 1PE-PL emission spectra to that of a known reference standard, RhB in anhydrous ethanol (with $\phi_{PL}$ = 0.49).\cite{casey} The quantum yield was calculated by integrating the emission spectra at 5 different concentrations. These are plotted against absorbance at the excitation wavelength for both the nc-SiQDs and RhB, shown in Figure \ref{fig:siqdotherchar}(c)-(d). The gradient of the trendline of luminescence intensity \textit{vs.} absorbance for both nc-SiQDs and RhB are used to compute the quantum yields of the nc-SiQDs by $\phi_{PL}^s = \phi_{PL}^r (m_s / m_r) (n_s / n_r)^2$, where ($\phi_{PL}^s, \phi_{PL}^r)$ is the (sample, reference) quantum yield, $(m_s, m_r)$ is the gradient of integrated PL \textit{vs.} absorbance of the (sample, reference),  $n_s = 1.496$ is the refractive index of the sample solvent (toluene), and $n_r = 1.365$ is the refractive index of the reference solvent (ethanol).\cite{kedenburg}

\subsubsection{Molar Absorptivities}

The molar absorptivities $\varepsilon(\omega)$ of the nc-SiQD samples were calculated by measuring the absorbances $ A(\omega)$ of nc-SiQD solutions at variable concentrations and applying the Beer-Lambert Law, $A(\omega) = \varepsilon (\omega) L C$, where $L$ is the optical path length (1 cm) and $C$ is the nc-SiQD concentration. The molar concentrations of the nc-SiQD samples were determined by taking the average diameter as measured from TEM imagery and calculating the ligands per nanocrystal with the mass fraction measured with TGA. The molar absorptivities of the nc-SiQD samples are shown in Figure \ref{fig:siqdotherchar}(e) and that of RhB from literature\cite{du} in Figure \ref{fig:siqdotherchar}(f). The molar absorptivities were used to determine the concentration of the samples in the 2PE-PL experiment by measuring the absorbance spectra of the samples with a spectrophotometer (PerkinElmer, Inc. Model  Spectrum 400) and fitting to an empirical model for the molar absorptivity to the Beer-Lambert Law. The molar concentration $C$ and number density $n_0$ of the samples are tabulated in the Supporting Information. Refer to the Supporting Information for additional details on these calculations.

\subsection{2PE-PL Measurements}

\subsubsection{Laser Source}

We two-photon excited PL using laser pulses from an OPA (Light Conversion Co. Model TOPAS-C, tuning range $240 < \lambda < 2600$ nm) pumped by a 1 kHz titanium-doped sapphire regenerative amplifier (Coherent, Inc. Model Libra HE USP). Excitation pulses passed through spectral, spatial, and polarization filters tailored for each excitation wavelength to ensure well-defined pulse spectra, Gaussian transverse beam profiles, and linear polarization throughout the tuning range. 

\subsubsection{Spatiotemporal Profile Characterization}

The pulse duration was measured by performing a SHG two-beam second-order autocorrelation using a barium borate crystal placed in the sample position.\cite{furey2} The average power was measured with a Si photodiode (PD) head (Coherent Model FieldMate, 650 - 1100nm) or a thermal head (Coherent Model FieldMate, 1200 - 2000 nm) and used to calibrate the reference photodiode detector.  The beam radius at the sample position was measured using an automated knife-edge technique. See the Supporting Information for details on characterization of the pulse duration and characterization of the beam radius.

\subsubsection{Experimental Setup}

A continuously variable neutral density (ND) filter wheel controls the incident intensity. A beam splitter reflects 4\% to a reference photodiode detector (Thorlabs, Inc. Model DET100A for 325 -- 1100 nm, Thorlabs PDA30G for 1200 -- 1400 nm) to monitor the incident laser power. The output of the photodiode detector was integrated and held for each pulse until the next trigger by a gated integrator (Stanford Research Systems Co. Model SR250).\cite{furey2} A plano-convex lens (with focal length $f = 500$ mm) focuses the transmitted beam into the sample which is contained in a 1 mm path length optical glass cuvette (Hellma GmbH \& Co. KG Model Z802689). A collection lens and fiber coupler injects a portion of the emitted PL into a 0.22 NA, $250 < \lambda < 1200$ nm, $d = 200\ \mu$m core, double clad, multimode optical fiber patch cable (Thorlabs Model FG200UCC). The spectra is then analyzed using a compact CCD spectrometer (Thorlabs Model CCS200) and recorded by computer. See the Supporting Information for further details on the experimental setup.

\subsubsection{Samples}

The samples were placed in a cuvette holder and the PL spectra recorded as a function of incident pulse energy. Care was taken to ensure that the incident intensity was kept below the threshold for bubble formation in the colloidal sample, as bubbles of vaporized solvent significantly increase scattered light. 

\subsection{Simulations of Efficiency of 2PE-PL Bio-Imaging}

The efficiency of a detectable signal from 2PE-PL in biological imaging applications is simulated using Equation \ref{eqbioimagingcalc2}. Effective attenuation coefficients for biological tissues were obtained from Smith \textit{et al.}\cite{smith}, Friebel \textit{et al.}\cite{friebel}, and Bashkatov \textit{et al.}\cite{bashkatov} In this simulation, we model the PL spectrum of the samples as a Gaussian peak with a center position at $E_0$ and peak width $\Delta E_0$, unique to each sample. The simulated samples include nc-SiQDs from this work; CdSe and CdTe quantum dots from Pu \textit{et al.}\cite{pu}; and various molecular fluorophores in literature: RhB 2PA spectrum and $\phi_{PL}$ from Makarov \textit{et al.}\cite{makarov} and PL spectrum from this work; BODIPY, Cascade Blue, Coumarin 307, DAPI, DiI, Fluorescein, and Lucifer Yellow 2PA spectra and $\phi_{PL}$ from Xu \textit{et al.}\cite{xu} and corresponding PL spectra for Cascade Blue, DAPI, and Fluorescein from Shapiro\cite{shapiro}; PL spectrum for BODIPY from Schmitt \textit{et al.}\cite{schmitt}; PL spectrum for Coumarin 307 from Mannekutla \textit{et al.}\cite{mannekutla}; PL spectrum for DiI from The Molecular Probes Handbook\cite{molprobes}; and the PL spectrum for Lucifer Yellow from omlc.\cite{omlclucifer}

\begin{acknowledgement}

This research was funded by Robert A. Welch Foundation Grants F-1038 and F-1464, and partially supported by the National Science Foundation through the Center for Dynamics and Control of Materials; an NSF MRSEC under Cooperative Agreement No. DMR-1720595. B. Mendoza acknowledges support from Consejo Nacional de Ciencia y Tecnolog\'{i}a, M\'{e}xico (Grant No. A1-S-9410). The majority of experimental work was performed at the Laboratorio de \'{O}ptica Ultrarr\'{a}pida at Centro de Investigaciones en \'{O}ptica, A.C. in Le\'{o}n, M\'{e}xico. The authors thank E. No\'{e}-Arias (Centro de Investigaciones en \'{O}ptica) for data acquisition program development, M. Olmos-L\'{o}pez (Centro de Investigaciones en \'{O}ptica) for access to spectrophotometer, and J. Clifford (University of Texas at Austin) and the Centro de Investigaciones en \'{O}ptica Machine Shop for machining assistance and part fabrication.

\end{acknowledgement}

%%%%%%%%%%%%%%%%%%%%%%%%%%%%%%%%%%%%%%%%%%%%%%%%%%%%%%%%
%% The same is true for Supporting Information, which should use the suppinfo environment.
%%%%%%%%%%%%%%%%%%%%%%%%%%%%%%%%%%%%%%%%%%%%%%%%%%%%%%%%

\begin{suppinfo}

The following file is available free of charge.
\begin{itemize}
  \item Supporting Information: 2PE-PL experimental setup diagram, spatiotemporal profile characterization, details of sample preparation and characterization, comparison of PL spectra by excitation channel, data archive, and the relationship of 2PA cross section to the isotropic molecular third-order nonlinear optical susceptibility tensor
\end{itemize}

\end{suppinfo}

%%%%%%%%%%%%%%%%%%%%%%%%%%%%%%%%%%%%%%%%%%%%%%%%%%%%%%%%
%% The appropriate \bibliography command should be placed here.
%% Notice that the class file automatically sets \bibliographystyle and also names the section correctly.
%%%%%%%%%%%%%%%%%%%%%%%%%%%%%%%%%%%%%%%%%%%%%%%%%%%%%%%%

\bibliography{ms}

\end{document}

% APS Revtex 4-1 class title
\title{\Large \bfseries Supporting Information:\\
\vspace{12pt}
\large Two-Photon Excitation Spectroscopy of \\Silicon Quantum Dots and Ramifications for Bio-Imaging}
%%  AIP Revtex 4-1 class title
%\title{\Large \bfseries Supporting Information:\\
%\large Im\{$\mathbf{\chi^{(3)}}$\} spectra of $\mathbf{110}$-cut GaAs, GaP, and Si near the two-photon absorption band edge}

% AIP Revtex 4-1 class authors
\author{Brandon J. Furey}
\email{furey@utexas.edu}
\affiliation{Physics Department, University of Texas at Austin, 2515 Speedway, C1600, Austin, TX, USA 78712}
\author{Ben Stacy}
\affiliation{Chemical Engineering Department, University of Texas at Austin, 2515 Speedway, C1600, Austin, TX, USA 78712}
\author{Tushti Shah}
\affiliation{Chemical Engineering Department, University of Texas at Austin, 2515 Speedway, C1600, Austin, TX, USA 78712}
\author{Rodrigo M. Barba-Barba}
\affiliation{Centro de Investigaciones en \'{O}ptica, A.C., Loma del Bosque 115, Colonia Lomas del Campestre, Le\'{o}n, Gto., M\'{e}xico 37150}
\author{Ramon Carriles}
\email{ramon@cio.mx}
\affiliation{Centro de Investigaciones en \'{O}ptica, A.C., Loma del Bosque 115, Colonia Lomas del Campestre, Le\'{o}n, Gto., M\'{e}xico 37150}
\author{Alan Bernal}
\affiliation{Centro de Investigaciones en \'{O}ptica, A.C., Loma del Bosque 115, Colonia Lomas del Campestre, Le\'{o}n, Gto., M\'{e}xico 37150}
\author{Bernardo S. Mendoza}
\affiliation{Centro de Investigaciones en \'{O}ptica, A.C., Loma del Bosque 115, Colonia Lomas del Campestre, Le\'{o}n, Gto., M\'{e}xico 37150}
\author{Brian A. Korgel}
\email{korgel@che.utexas.edu}
\affiliation{Chemical Engineering Department, University of Texas at Austin, 2515 Speedway, C1600, Austin, TX, USA 78712}
\affiliation{Texas Materials Institute, University of Texas at Austin, 204 E. Dean Keeton St., C2201, Austin, TX USA 78712}
\author{Michael C. Downer}
\email{downer@physics.utexas.edu}
\affiliation{Physics Department, University of Texas at Austin, 2515 Speedway, C1600, Austin, TX, USA 78712}

\maketitle

% Disable subsubsections in the TOC
\makeatletter
\def\l@subsubsection#1#2{}
\makeatother
% Hyperlinks black in TOC and display TOC
{\hypersetup{linkcolor=black}\tableofcontents}

\clearpage

\FloatBarrier
\section{\label{sec:expsetup}2PE-PL Experimental Setup}

\begin{figure}[ht]
  \includegraphics[width=\linewidth]{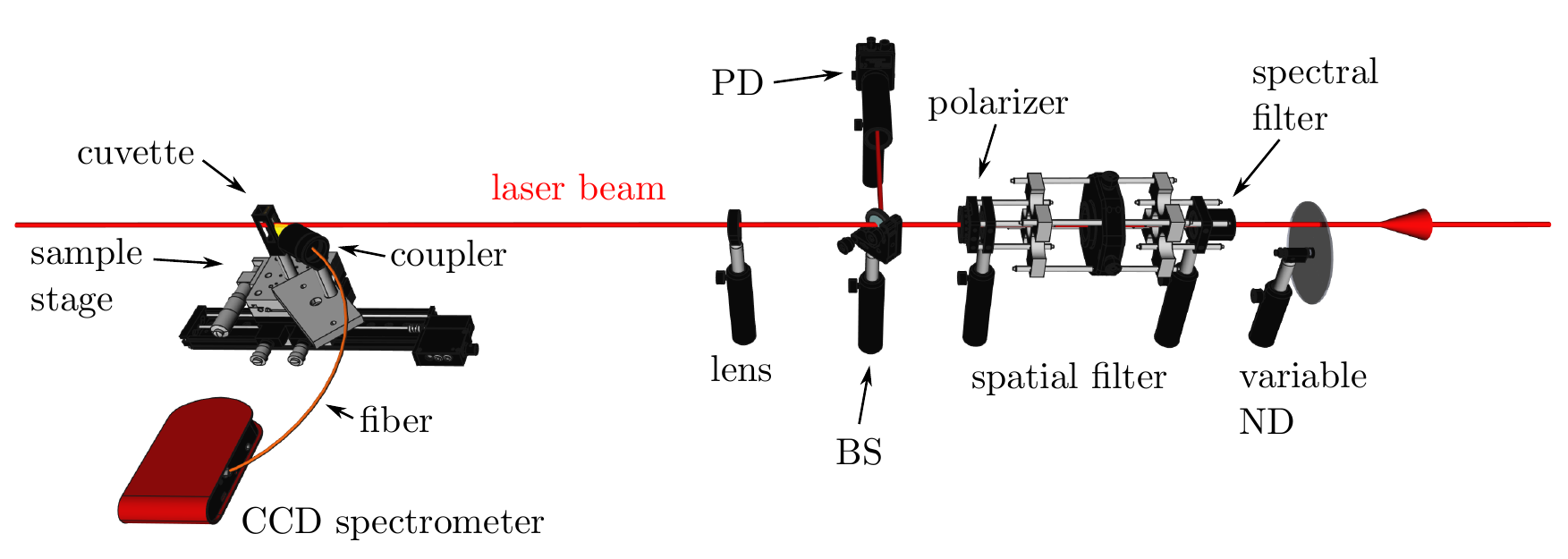}
  \caption{2PE-PL experimental setup. The beam passes through spectral and spatial filters and a thin-film polarizer. A beam splitter (BS) reflects 4\% to a reference photodiode detector (PD) to monitor the incident laser power. The transmitted beam is focused by a lens into a sample cuvette mounted in a cuvette holder on the sample stage. The fiber coupler collects a portion of the emitted 2PE-PL which is measured using a fiber spectrometer and recorded by computer. Image includes CAD models courtesy of Thorlabs, Inc., Zaber Technologies, Inc., and Edmund Optics, Inc. All rights reserved.}
  \label{fig:expsetup}
\end{figure}

Fig. \ref{fig:expsetup} shows the 2PE-PL experimental setup.

\FloatBarrier
\section{\label{sec:spatiotemporalprofiles}Spatiotemporal Profile Characterization}

Knowledge of the absolute spatiotemporal intensity profile is not required in a relative 2PE-PL measurement in the weak 2PA limit, \textit{i.e.}, when $\sigma n_0 z I_0 \ll 1$. However, the intensity dependence of higher-order terms in the expansion of Eq. 5 in the article are different than the lowest-order term in that they are not proportional to $f_{det}$. Thus the intensity must be characterized for cases when higher-order terms are included in the sum. A Gaussian spatiotemporal pulse profile is assumed of the form 
\begin{equation}
\label{eqintensity}
I(0,r,t) = I_0 e^{-2 r^2 / w^2} e^{-2t^2 / \tau_g^2},
\end{equation}
with on-axis peak intensity given by
\begin{equation}
\label{eqi0}
I_0 = \bigg( \frac{2}{\pi}\bigg)^{3/2} \frac{\varepsilon_P}{w^2 \tau_g},
\end{equation}
$w$ is the beam radius, $\tau_g = \tau_{FWHM}/\sqrt{2 \mathrm{ln}2}$ is the Gaussian pulse duration parameter expressed in terms of the FWHM pulse duration, $\varepsilon_P \rightarrow \varepsilon_P (1 - R)$ is the incident pulse energy, and $R$ is the Fresnel power reflection coefficient. 

\subsection{\label{sec:spatiotemporalprofilesbeamradiii}Beam Radii}

\begin{figure}
  \includegraphics[scale=1.0]{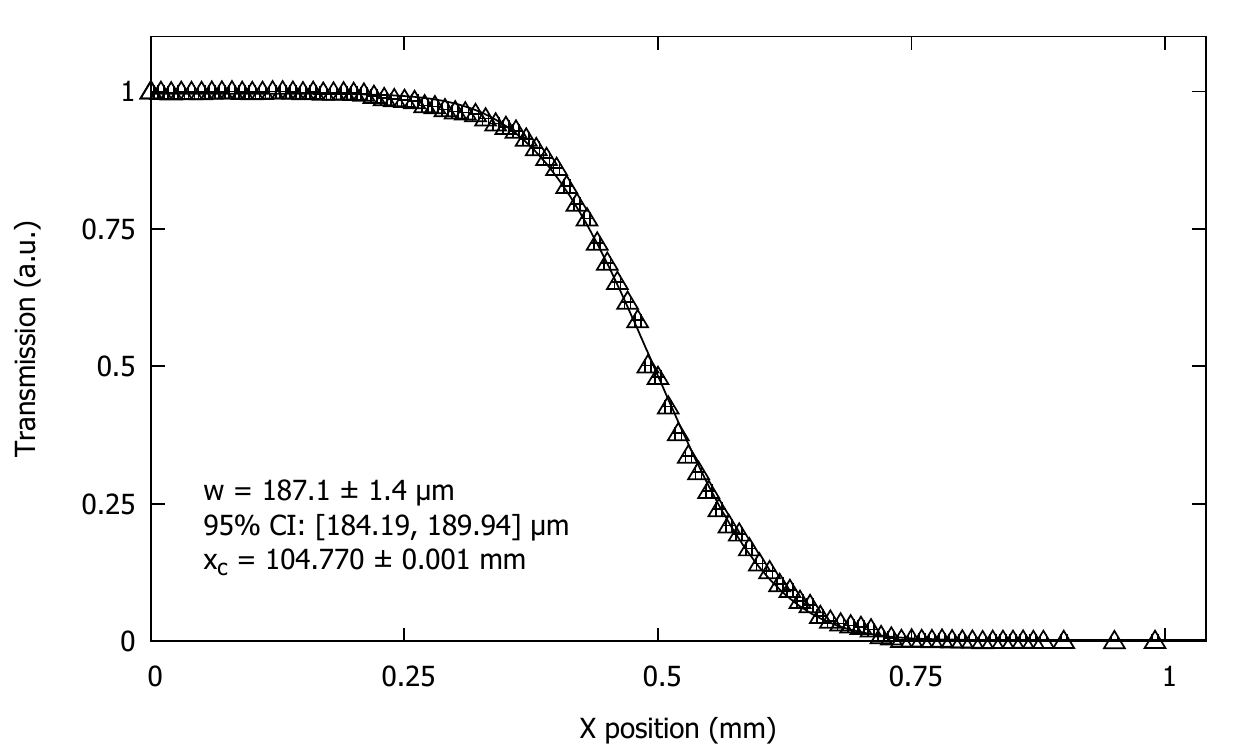}
  \caption{Knife-edge transect to characterize transverse beam profile at sample location at $\lambda = 800$ nm.}
  \label{fig:knifeedge}
\end{figure}

We present here the beam radii as measured using a knife-edge transect for each wavelength studied. The transmitted intensity during a knife-edge transect is 
\begin{equation}
\label{eqknifeedge}
T(\Delta x) = \frac{a}{2} \Bigg\{ 1 - \mathrm{erf}\bigg[\frac{\sqrt{2}\Delta x}{w(Z)}\bigg]\Bigg\} + b,
\end{equation}
where $a$ is the amplitude, $b$ is an intensity offset, and $\Delta x = x - x_c$ is the transverse position of the knife edge relative to the beam center, $x_c$. An example of this analysis is shown in Fig. \ref{fig:knifeedge} and the values for all wavelengths studied in this work are tabulated in Tab. \ref{tab:beamradii}.

\begin{table}[ht]
\centering
 \caption{\label{tab:beamradii}Summary of beam radii at sample location.}
  \begin{tabular}{c c c c}
  \hline
  $\lambda$ (nm)& $\hbar \omega$ (eV)& $w(Z)$ ($\mu$m)& $\delta w(Z)$ ($\mu$m)\\
  \hline
  650 & 1.91 & 150 & 4\\
  700 & 1.77 & 120 & 6\\
  750 & 1.65 & 90 & 6\\
  800 & 1.55 & 187.1 & 1.4\\
  850 & 1.46 & 197 & 13\\
  1250 & 0.99 & 192 & 4\\  
  \hline
  \end{tabular}
\end{table}

\FloatBarrier
\subsection{\label{sec:spatiotemporalprofilespulseduration}Pulse Durations}

\begin{figure}[ht]
  \includegraphics[scale=1.0]{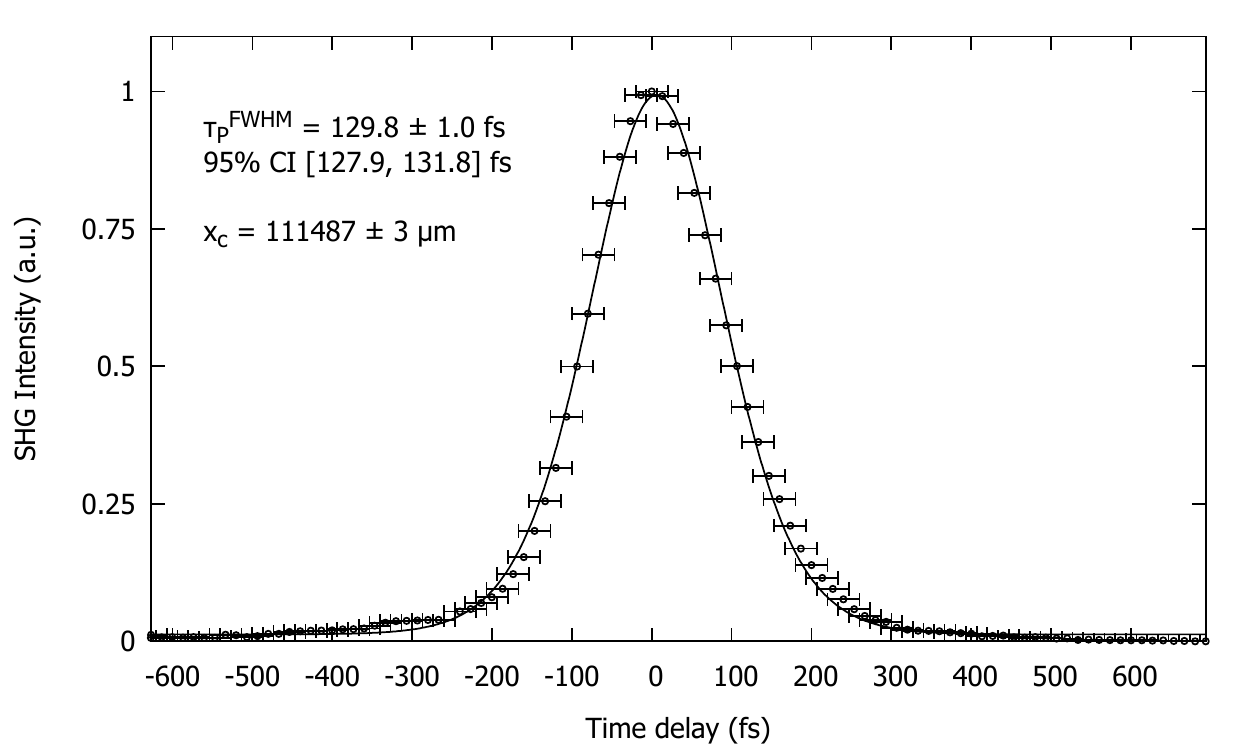}
  \caption{Noncollinear two-beam SHG second-order autocorrelation at $\lambda = 800$ nm using a barium borate nonlinear crystal (black circles with error bars) and fit (black curve).}
  \label{fig:shgpulseduration}
\end{figure}

The pulse duration was measured using a second-order autocorrelation technique, where a second beam was intersected with the primary beam at an angle of $\psi \approx 15^{\circ}$. A barium borate nonlinear crystal was placed at the overlap position, and second-harmonic generation (SHG) occurs along the bisector of the two beams when they are temporally and spatially overlapped and phase-matching conditions are satisfied. Adjusting the relative time delay between the two pulses $\Delta t = t - t_0$, using a time delay stage in the primary beam path, produces a SHG peak of the form
\begin{equation}
\label{eqshgautocorrelator}
T^{SHG}(\Delta t) = \frac{3 a }{\mathrm{sinh}^2(2.7196 \Delta t/\tau_a)} \Bigg[ \frac{2.7196 \Delta t}{\tau_a} \mathrm{coth} \bigg(\frac{2.7196 \Delta t}{\tau_a} - 1 \bigg) \Bigg] + b,
\end{equation}
where $a$ is an amplitude parameter, $b$ is an intensity offset parameter, $\Delta t = t - t_0$ is the time delay, measured time $t$ from the peak center position parameter $t_0$, and $\tau_a = 1.54 \tau_{FWHM}$ is the autocorrelation width parameter. The autocorrelation width parameter is related to the full-width at half-maximum (FWHM) pulse duration for a hyperbolic secant squared pulse by a factor of 1.54. In this article, a Gaussian profile was assumed, for which the corresponding Gaussian pulse duration parameter is related to the FWHM pulse duration by $\tau_g = \tau_{FWHM}/\sqrt{2 \mathrm{ln}2}$. Five measurements were performed at each wavelength. An example of this analysis is shown in Fig. \ref{fig:shgpulseduration} and the values for all wavelengths studied in this work are tabulated in Tab. \ref{tab:pulsedur}.

\begin{table}[ht]
\centering
 \caption{\label{tab:pulsedur}Summary of pulse durations at sample location.}
  \begin{tabular}{c c c c}
  \hline
  $\lambda$ (nm)& $\hbar \omega$ (eV)& $\tau_{FWHM}$ (fs)& $\delta \tau_{FWHM}$ (fs)\\
  \hline
  650 & 1.91 & 154 & 6\\
  700 & 1.77 & 123 & 3\\
  750 & 1.65 & 123 & 3\\
  800 & 1.55 & 125 & 2\\
  850 & 1.46 & 148 & 5\\
  1250 & 0.99 & 104 & 4\\  
  \hline
  \end{tabular}
\end{table}

\FloatBarrier
\section{\label{sec:sampleprep}Sample Preparation}

\subsection{\label{sec:sampleprep:materials}Materials}

Nanocrystal samples were prepared from the following starting materials: Commercial flowable oxide polymer FOx-16 (Dow Corning Corp.) containing 16\% hydrogen silsesquioxane (HSQ) by weight in methyl isobutyl ketone; ethanol (MilliporeSigma Co., 99\%); methanol (MilliporeSigma, 99\%); hydrofluoric acid (MilliporeSigma, 48\%); 1- dodecene (MilliporeSigma, 95\%); toluene (Fisher Scientific International, Inc., 99.9\%); chloroform (Acros Organics Co., 99.8\%); hydrochloric acid (Fisher, 25\%); Rhodamine 101 (MilliporeSigma, 100\%); and Rhodamine B (MilliporeSigma, 100\%) were purchased and used as received.

\subsection{\label{sec:sampleprep:hydrideterminated}Hydride-Termination of nc-SiQDs}

\begin{figure}
  \includegraphics[width=0.3\textwidth]{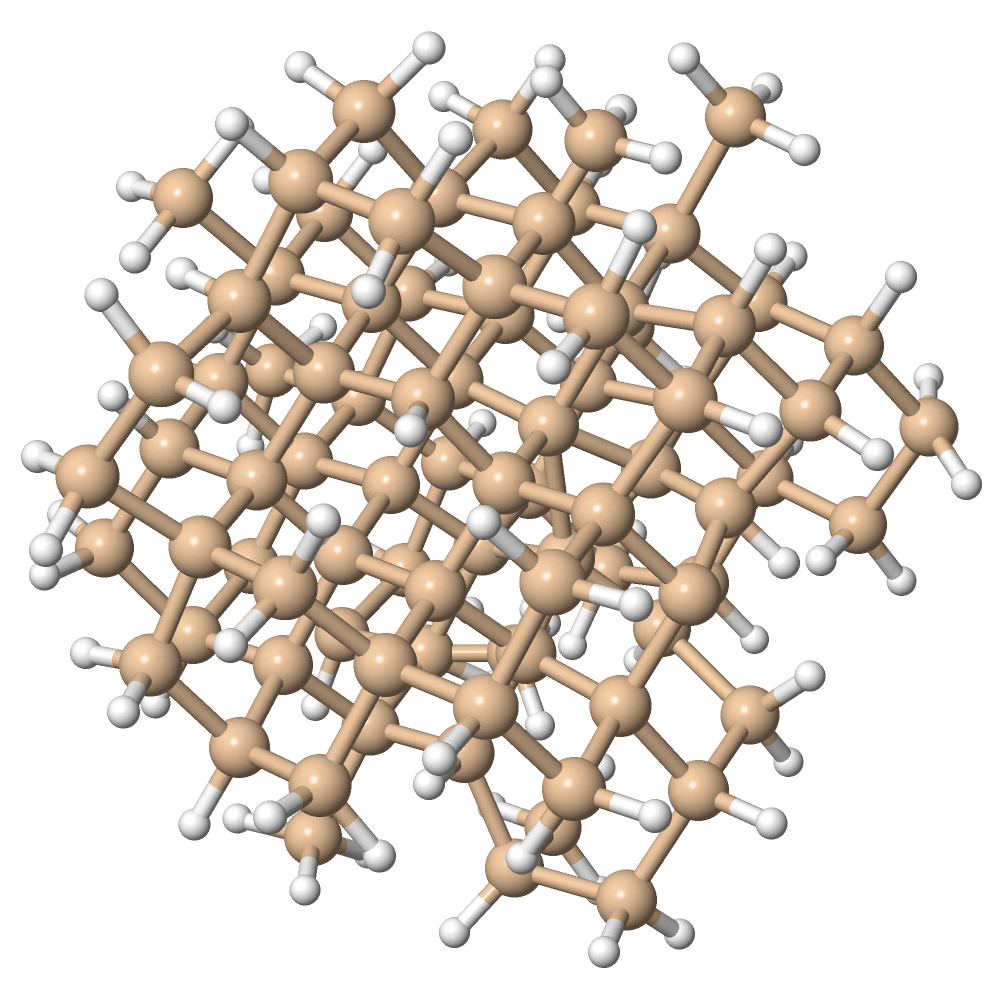}
  \caption{Model of nc-SiQD with $d \approx 1.8$ nm passivated with hydrogen.\cite{vmols}}
  \label{fig:ncsiqddiagram}
\end{figure}

Hydrogen-passivated nanocrystals were synthesized as described previously.\cite{hessel} Briefly, 4 g of HSQ was heated from room temperature to 1000--1100$^{\circ}$C over one hour and held at the set temperature for one hour under 95\% N2, 5\% H2 forming gas flow. The brown, glassy product produced was ground with an agate mortar and pestle until it is a light-brown powder and shaken in a wrist-action shaker for 9 h with 3 mm borosilicate glass beads to further reduce grain size. 300 mg of the ground powder was etched in 1 mL of 25\% HCl and 10 mL of 48\% HF in the dark for 2--3 h. Nanocrystals were isolated by centrifugation at 8000 rpm for 5 min, rinsed twice with ethanol, and once with chloroform. This procedure yields hydride-terminated nc-SiQDs, such as the the one modeled in Fig. \ref{fig:ncsiqddiagram}.

\subsection{\label{sec:sampleprep:alkenepassivation}Alkene Surface Passivation of nc-SiQDs}

After centrifugation, the chloroform supernatant is discarded, and the light brown precipitate is dispersed in 8 mL of 1-dodecene. The resulting brown, turbid dispersion is transferred to a three-neck round-bottom flask and degassed with 3 freeze--pump--thaw cycles on a greaseless Schlenk line. The dispersion is heated at 190$^{\circ}$C for 20 h. After about 30 min of heating, the turbid brown dispersion turns to an optically clear orange dispersion, indicating passivation of the nc-SiQDs. The alkene passivated nc-SiQDs are washed three times by precipitation with ethanol as an antisolvent (15 mL) and redispersed in 1 mL of toluene for use in experiments.

\FloatBarrier
\section{\label{sec:samplechar}Sample Characterization}

\subsection{\label{sec:samplechar:tem}Transmission Electron Microscopy}

Average nanocrystal diameters and size distributions were determined by imaging individual nanocrystals with a high-resolution transmission electron microscope (JEOL Ltd. Model 2010F HR-TEM) operated at 200 kV. Refer to Figs. 1(a)-(e) in the article. Nanocrystals were drop-cast onto graphene TEM substrates prepared on lacey carbon-coated copper grids (Electron Microscopy Sciences, Inc.) by dropping 3 $\mu$L of ethanol dispersion on graphene (0.1 mg/mL). The nc-SiQD samples annealed at $1000^{\circ}$C had a size distribution of $1.8 \pm 0.2$ nm and at $1100^{\circ}$C had a size distribution of $2.3 \pm 0.3$ nm. 

The high resolution images in Figs. 1(c)-(d) in the article were obtained with aberration-corrected scanning TEM (acSTEM) using a JEOL NEOARM TEM with an 80 kV accelerating voltage and a point-to-point STEM resolution of 0.11 nm. Graphene-enhanced lacey carbon TEM grids were purchased from Electron Microscopy Sciences (Cat. no. GF1201). A dilute nanocrystal solution in chloroform was dropcast onto the grid and stored in a vacuum chamber overnight before imaging.

\subsection{\label{sec:samplechar:xrd}X-Ray Diffraction}

XRD was performed on each of the nc-SiQD samples using a Rigaku R-Axis Spider diffractometer using Cu K$\alpha$ radiation ($\lambda = 0.15418$ nm). The quantum dots were deposited on a glass slide, the solvent was evaporated, and then the powder was placed on a nylon loop. Two-dimensional diffraction data were collected for 10 min while rotating the sample stage at $10^{\circ}$ per minute. 2D diffraction data were radially integrated with 2DP software and are shown in Fig. 1(f) in the article. The nc-SiQDs are confirmed to be crystalline by the presence of (111), (220), and (311) diffraction peaks for each of these crystal planes in $m3m$ crystalline Si.

\subsection{\label{sec:samplechar:tga}Thermal Gravimetric Analysis}

Thermal gravimetric analysis (TGA) was performed using an automated ultra-micro balance (Mettler-Toledo International, Inc. Model TGA-1), adding 5 mg of passivated nc-SiQDs to a 70 $\mu$L alumina crucible (Mettler Toledo). Samples were heated at a rate of 20$^{\circ}$C/min from 40$^{\circ}$C to 800$^{\circ}$C. The sample was held at 100$^{\circ}$C for 30 minutes to evaporate residual solvents and at 800$^{\circ}$C for 30 min to ensure all ligand had evaporated. Measurements were conducted under 50 mL/min of air flow. The relative mass of Si core to 1-dodecene in the passivated samples were determined from the remaining mass of Si and the total weight loss due to the removal of the ligands. The mass fraction temperature curves are shown in Fig. 2(a) in the article.

\subsection{\label{sec:samplechar:plpleabs}PL, PLE, and Absorbance Spectra}

One-photon excited PL (1PE-PL) and photoluminescence excitation (PLE) spectra in the ultraviolet--visible (UV--Vis) wavelength range were acquired on a fluorescence spectrophotometer (Varian, Inc. Model Cary Eclipse). UV--Vis absorbance spectroscopy was performed on a UV--Vis spectrophotometer (Varian, Inc. Model Cary 50 Bio UV--Vis). PL quantum yields were estimated relative to RhB in anhydrous ethanol (see following section on quantum yield calculations).

\subsection{\label{sec:methods:samplechar:qy}Quantum Yield Calculations}

The quantum yield ($\phi_{PL}$) was determined by comparing the integrated emission spectra of the nc-SiQDs to that of a known reference standard, RhB with $\phi_{PL}$ = 0.49.\cite{casey} The quantum yield was calculated by integrating the emission spectra at 5 different concentrations. These are plotted against absorbance at the excitation wavelength for both the nc-SiQDs and RhB, shown in Fig. 2(c)-(d) in the article. The gradient of the trendline of luminescence intensity vs. absorbance for both nc-SiQDs and RhB are used in the equation in the Sample Characterization subsection of the Methods section in the article, where ($\phi_{PL}^s, \phi_{PL}^r)$ is the (sample, reference) quantum yield, $(m_s, m_r)$ is the gradient of integrated PL vs. absorbance of the (sample, reference),  $n_s = 1.496$ is the refractive index of the sample solvent (toluene), and $n_r = 1.365$ is the refractive index of the reference solvent (ethanol).\cite{kedenburg} The calculated quantum yields for the nc-SiQDs in this study are $\phi_{PL} = 0.064$ for $d = 1.8 \pm 0.2$ nm and $\phi_{PL} = 0.060$ for $d=2.3 \pm 0.3$ nm. 

\subsection{\label{sec:samplechar:molarabsorptivity}Molar Absorptivity and Concentration Calculations}

The molar absorptivity of the nc-SiQD samples were determined by first measuring the absorbance spectra at varying mass concentrations and then calculating the conversion factor between mass concentration and molar concentration. Empirical functions were used to model the molar absorptivity spectra, and then fit to the absorbance spectra of the samples at the time of the 2PE-PL experiment using the Beer-Lambert Law. This allowed determination of the number density of the samples. The calculations are detailed in the following section.

\FloatBarrier
\subsubsection{\label{sec:samplechar:molarabsorptivity:absmass}Absorbance Spectra at Varying Mass Concentrations}

\begin{figure}
  \includegraphics[scale=0.4]{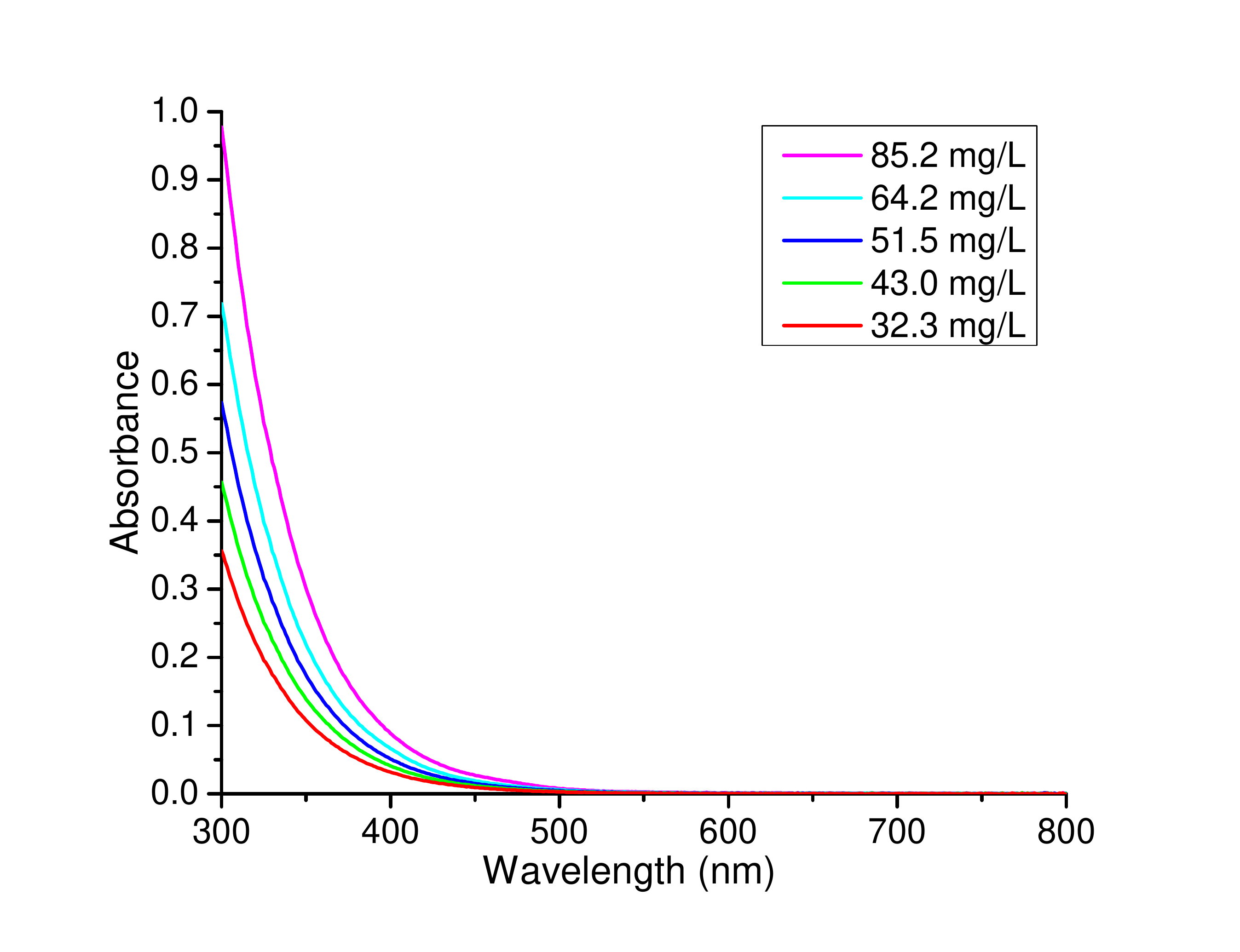}
  \caption{Absorbance spectra of dodecene-passivated nc-SiQDs annealed at 1000$^{\circ}$C ($d = 1.8 \pm 0.2$ nm) at varying mass concentrations.}
  \label{fig:absmass1000}
\end{figure}

\begin{figure}
  \includegraphics[scale=0.4]{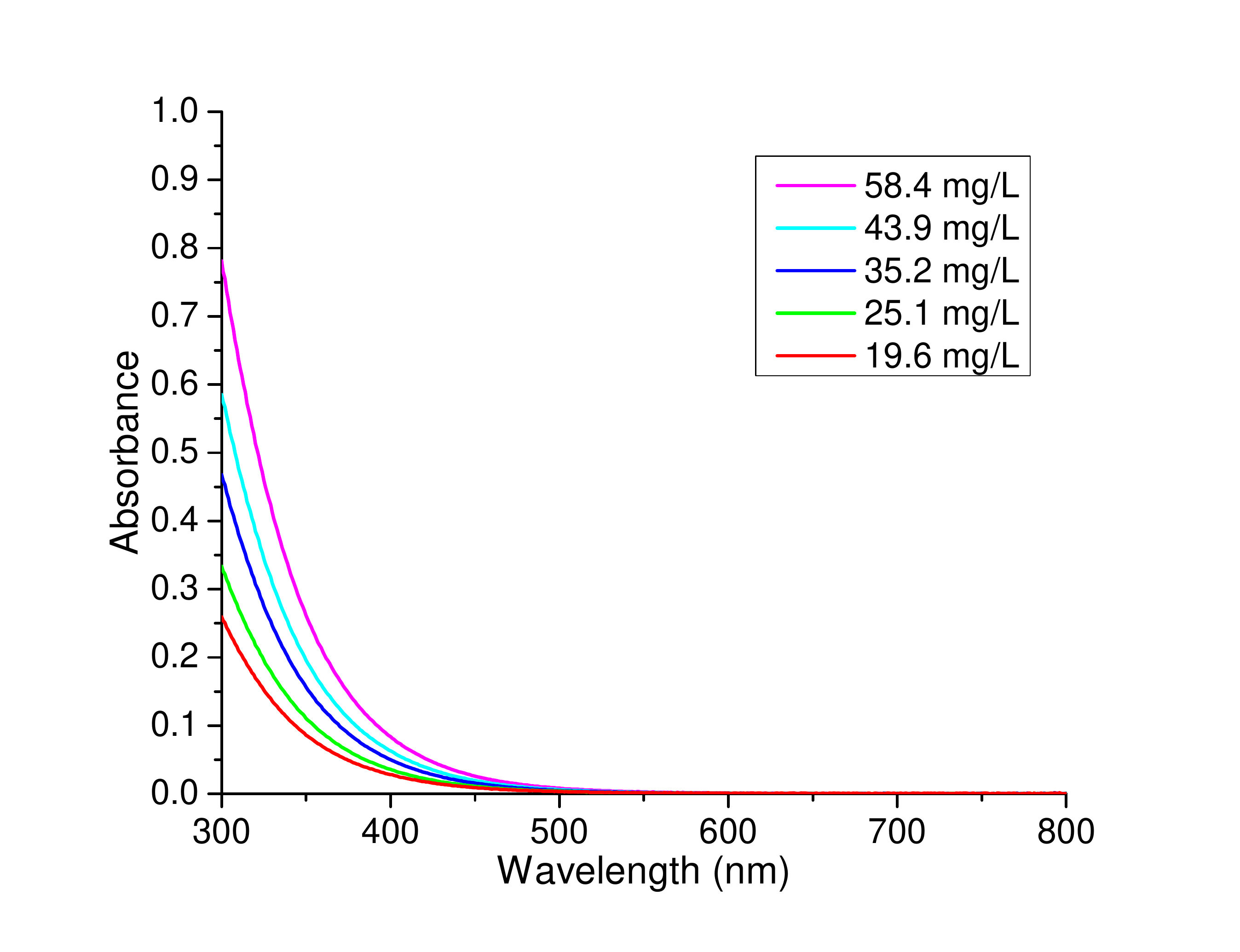}
  \caption{Absorbance spectra of dodecene-passivated nc-SiQDs annealed at 1100$^{\circ}$C ($d = 2.3 \pm 0.3$ nm) at varying mass concentrations.}
  \label{fig:absmass1100}
\end{figure}

The absorbance spectra of the nc-SiQD samples in a 1 cm path length cuvette were measured at varying mass concentrations using a UV--Vis spectrophotometer (Varian, Inc. Model Cary 50 Bio UV--Vis). Refer to Figs. \ref{fig:absmass1000} and \ref{fig:absmass1100}.

\FloatBarrier
\subsubsection{\label{sec:samplechar:molarabsorptivity:molweight}Molecular Weight Calculations}

The molecular weights (MW) of the nc-SiQD samples were calculated using the average nanocrystal diameter $d$ determined by HR-TEM imagery. The nanocrystals are assumed to be spherical and the volume is computed. The total volume of a nanocrystal is divided by the volume of a silicon diamond cubic unit cell. Each unit cell contains 8 atoms. The number of silicon atoms in the core was converted to a molecular weight, and the mass fraction of ligand to core as determined by TGA analysis was used to estimate the total molecular weight for each nc-SiQD sample size. These calculations are detailed in Tab. \ref{tab:molweight}.

\begin{table}[ht]
\centering
 \caption{\label{tab:molweight}Molecular weight calculations.}
 \resizebox{\textwidth}{!}{%
  \begin{tabular}{c c c c c c c c}
  \hline
  $d$ (nm)& Volume (nm$^3$)& \# unit cells/nc-SiQD& \# Si atoms/nc-SiQD& Core MW (g/mol)& Ligand mass fraction& Total MW (g/mol)\\
 \hline
 1.8& 3.0& 19& 152& 4284& 0.517& 8870\\
 2.3& 6.3& 39& 318& 8938& 0.435& 15820\\ 
  \hline
  \end{tabular}
  }
\end{table}

\FloatBarrier
\subsubsection{\label{sec:samplechar:molarabsorptivity:molabs}Molar Absorptivity}

\begin{figure}
  \includegraphics[scale=0.4]{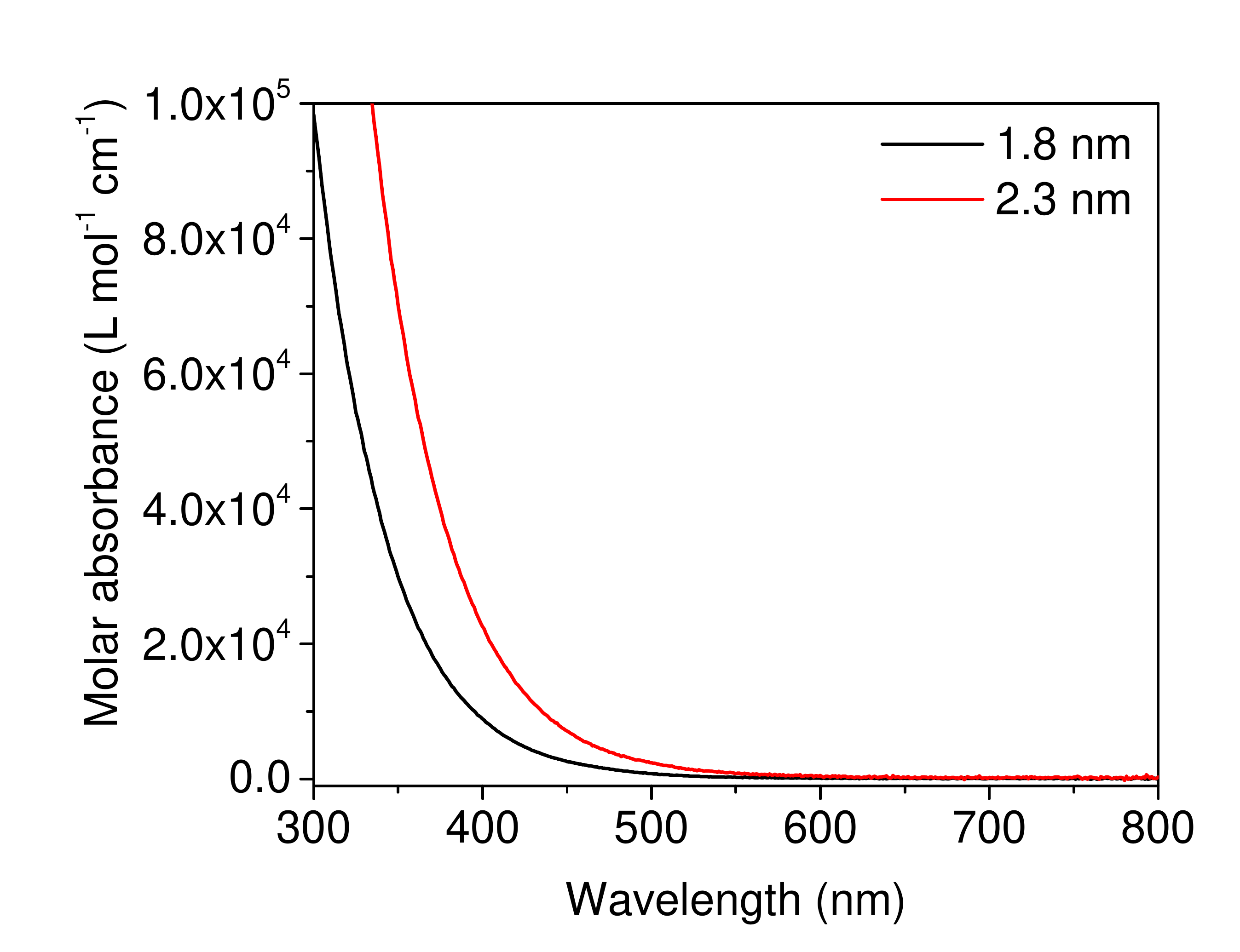}
  \caption{Molar absorptivity spectra of dodecene-passivated nc-SiQD samples with $d = 1.8 \pm 0.2$ and $d = 2.3 \pm 0.3$ nm.}
  \label{fig:suppmolabs}
\end{figure}

The molar absorptivity $\varepsilon(\lambda)$ of the nc-SiQD samples were calculated according to the Beer-Lambert Law
\begin{equation}
\label{eqbeerslaw}
A(\lambda) = \varepsilon(\lambda) L C,
\end{equation}
where $A$ is the sample absorbance, $L$ is the path length, and $C$ is the sample concentration. This calculation was repeated for each of the 5 different concentrations of each sample measured, and the values of the molar absorptivity obtained were averaged to give the final molar absorptivity spectra, shown in Fig. \ref{fig:suppmolabs} and in Fig. 2(e) in the article.

\FloatBarrier
\subsubsection{\label{sec:samplechar:molarabsorptivity:concentration}Concentration and Number Density}

An exponential decay function was used to model the nc-SiQD molar absorptivity spectra, of the form
\begin{equation}
\label{eqexpdecay}
\varepsilon(\lambda) = A e^{-\lambda/\Lambda},
\end{equation}
with $A$ an amplitude and $\Lambda$ a decay constant. The fit parameters are tabulated in Tab. \ref{tab:ncsiqdmolabsfitparam}.

\begin{table}[ht]
\centering
 \caption{\label{tab:ncsiqdmolabsfitparam}Fit parameters for molar absorptivities of nc-SiQD samples.}
  \begin{tabular}{c c c}
  \hline
  $d$ (nm)& $A$ (L mol$^{-1}$ cm$^{-1}$)& $\Lambda$ (nm)\\
  \hline
1.8& 1.31117E+08& 41.74077\\
2.3& 1.7312E+08& 44.81807\\
  \hline
  \end{tabular}
\end{table}

A composition of three Gaussian peaks was used to model the RhB molar absorptivity spectra, obtained from Du \textit{et al.},\cite{du} of the form
\begin{equation}
\label{eq3gaussians}
\varepsilon(\lambda) = A_1 e^{-(\lambda - \lambda_1)^2/\Lambda_1^2} + A_2 e^{-(\lambda - \lambda_2)^2/\Lambda_2^2} + A_3 e^{-(\lambda - \lambda_3)^2/\Lambda_3^2},
\end{equation}
where $A_i$ are the peak amplitudes, $\lambda_i$ are the peak center positions, and $\Lambda_i$ are the peak widths. The fit parameters are tabulated in Tab. \ref{tab:rhbmolabsfitparam}.

\begin{table}[ht]
\centering
 \caption{\label{tab:rhbmolabsfitparam}Fit parameters for molar absorptivities of RhB.}
  \begin{tabular}{l | l}
  \hline
  Parameter& Value\\
  \hline
$A_1$ (L mol$^{-1}$ cm$^{-1}$)& 5.53233E+06\\
$\lambda_1$ (nm)& 0.0\\
$\Lambda_1$ (nm)& 144.30268\\
$A_2$ (L mol$^{-1}$ cm$^{-1}$)& 3.757723547E+04\\
$\lambda_2$ (nm)& 519.64888\\
$\Lambda_2$ (nm)& 31.65463\\
$A_3$ (L mol$^{-1}$ cm$^{-1}$)& 8.42270341E+04\\
$\lambda_3$ (nm)& 544.21958\\
$\Lambda_3$ (nm)& 16.7233\\
  \hline
  \end{tabular}
\end{table}

These functions are then fit using Eq. \ref{eqbeerslaw} to the absorbance spectra of the samples at the time of the 2PE-PL experiment, determining the molar concentration $C$ and number density $n_0$. The path length of the cuvette was $L = 1.00$ mm. These results are tabulated in Tab. \ref{tab:concnumbdens} below.

\begin{table}[ht]
\centering
 \caption{\label{tab:concnumbdens}Molar concentration and number density of samples.}
  \begin{tabular}{l l c c c c}
  \hline
 Sample& $\lambda$ (nm)& $C$ (M)& $\delta C$ (M)& $n_0$ (cm$^{-3}$)& $\delta n_0$ (cm$^{-3}$)\\
  \hline
nc-SiQD $d = 1.8$ nm& 650, 700, 750& 6.5348E-03& 2.3657E-05& 3.9353E+18& 1.4247E+16\\
nc-SiQD $d = 1.8$ nm& 800, 850, 1250& 6.2111E-03& 3.6816E-05& 3.7404E+18& 2.2171E+16\\
nc-SiQD $d = 2.3$ nm& 650, 700, 750& 2.2867E-03& 1.1296E-05& 1.3771E+18& 6.8026E+15\\
nc-SiQD $d = 2.3$ nm& 800, 850, 1250& 2.2644E-03& 1.2946E-05& 1.3637E+18& 7.7963E+15\\
RhB& 650, 700, 750& 3.0338E-04& 5.4619E-06& 1.8270E+17& 3.2893E+15\\
RhB& 800, 850, 1250& 3.0809E-04& 5.4986E-06& 1.8554E+17& 3.3113E+15\\
  \hline
  \end{tabular}
\end{table}

\FloatBarrier
\section{\label{sec:excitationchannel}Comparison of PL Spectra by Excitation Channel}

\begin{figure}
  \includegraphics[scale=1.0]{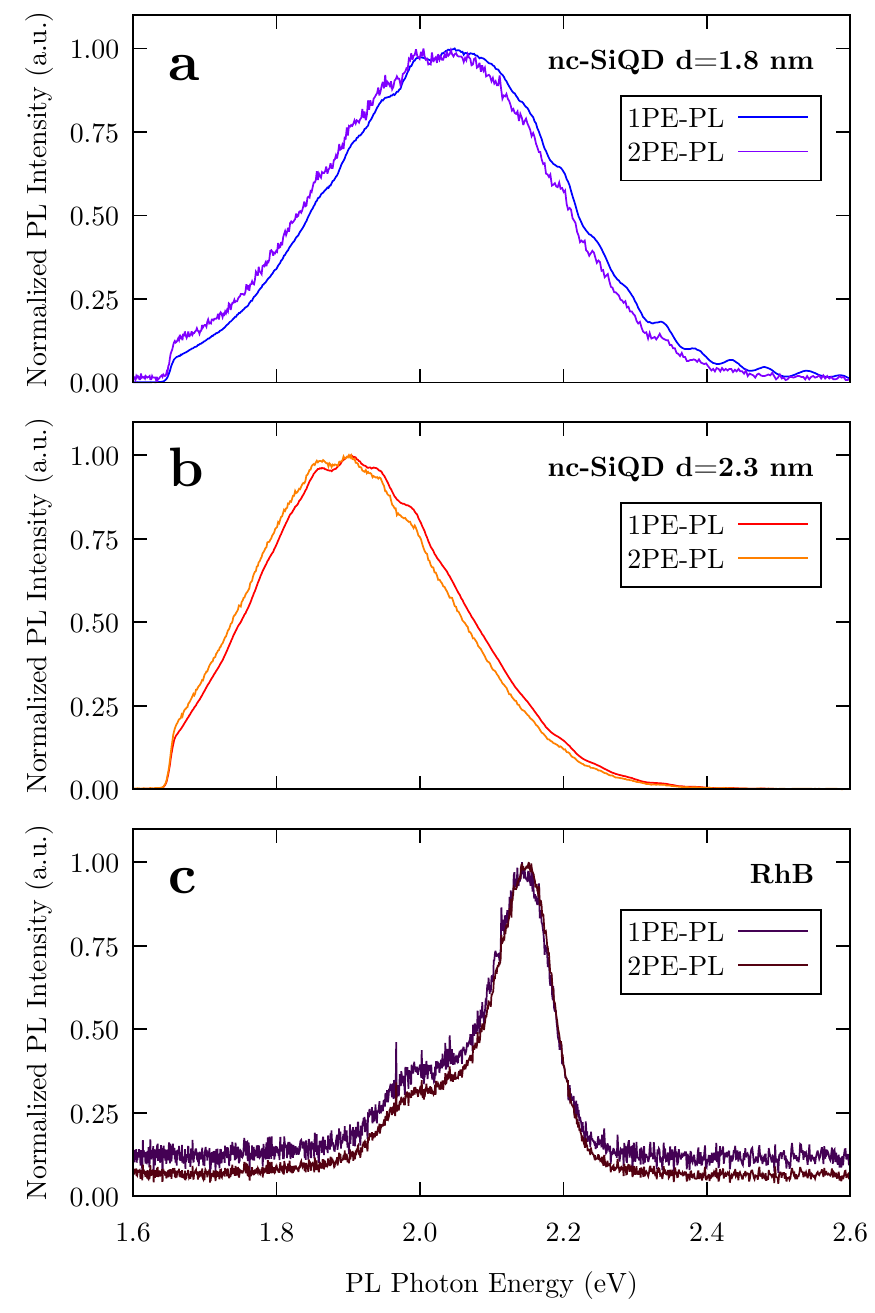}
  \caption{Normalized PL spectra for nc-SiQD samples with $d = 1.8$ nm (a), $d = 2.3$ nm (b), and RhB (c). The excitation channels are indicated; 1PE-PL is excited with low-power $\lambda = 400$ nm SHG and 2PE-PL is excited with high-power $\lambda = 800$ nm amplified pulses. The laser source is a titanium-doped sapphire regenerative amplifier $\big[$Spectra-Physics, Inc. Model Spitfire for (a) and (b), and Coherent, Inc. Model Libra HE USP for (c)$\big]$. The detector system includes a fiber coupled to a spectrometer $\big[$Horiba, Ltd. Model iHR 320 for (a) and (b), and Thorlabs Model CCS200 for (c)$\big]$.}
  \label{fig:suppchannels}
\end{figure}

The quantum yield $\phi_{PL}$ for the samples can be assumed to be the same for 2PE as 1PE so long as the PL emission spectra between the two excitation channels do not appreciably differ.\cite{diener,rumi} It has been observed that at room temperature, the PL spectra and PL quantum yields are nearly the same for both excitation channels in nc-SiQDs.\cite{diener} The small shoulders observed in the 2PE-PL spectra using the Thorlabs, Inc. CCS200 fiber-coupled CCD spectrometer compared to the 1PE-PL spectra observed using the Varian, Inc. Model Cary 50 Bio UV--Vis spectrophotometer were due to modulations caused by an etalon formed by a glass window between the grating and the CCD.\cite{thorlabswebsite} This was compensated for with an amplitude correction procedure detailed in Sec. \ref{sec:expsetup:spectrometercorrection}.\cite{thorlabsmanual} The integrated PL photon number is insensitive to any remaining small modulations in the spectral structure. We compared the PL emission spectra from both 1PE and 2PE channels and found that they are indeed nearly identical. Refer to Fig. \ref{fig:suppchannels}.

\FloatBarrier
\section{\label{sec:dataarchive}Data Archive}

Plots of the 2PE-PL intensity scans used in the 2PE-PL analysis in this article are presented here along with tables of the fit parameters. The raw data that support the findings of this study are available from the corresponding authors upon reasonable request.

\subsection{\label{sec:expsetup:spectrometercorrection}Spectrometer Wavelength Calibration and Amplitude Correction}

The wavelength calibration of the Thorlabs CCS200 spectrometer used for collecting 2PE-PL data was performed using 8 lines from a NeAr lamp and 7 lines from a Hg lamp. The measured wavelengths were consistently offset from the true values by $\lambda_{offset} = -4.54 \pm 0.10$ nm. The measured data was then calibrated by $\lambda = \lambda_{measured} - \lambda_{offset}$. 

The Thorlabs CCS200 spectrometer has a glass window between the diffraction grating and CCD which forms an etalon and causes regular modulations in the observed spectrum.\cite{thorlabswebsite} These modulations have a peak-to-peak amplitude of $\approx 20$--$25\%$ and a period of $\approx 0.2$ eV. In addition, the Thorlabs CCS200 spectrometer is affected by CCD, diffraction grating, and fiber transmission efficiencies. These effects were compensated for with an amplitude correction\cite{thorlabsmanual}

\FloatBarrier
\subsection{\label{sec:expsetup:spectralineshapefit}2PE-PL Spectra Lineshape Fitting}

The PL spectra $S_{PL}(E)$ is proportional to the PL photon number within a photon energy bin of width $\mathrm{d}E$ centered at $E$. It can be empirically modeled as a composition of two Gaussian peaks and the laser lineshape $S_L(E)$ is modeled as a Lorentzian. The total detected spectra is then described by the sum of these spectral lineshapes,  
\begin{equation}
\label{eqdetectorlineshapes}
S_{tot} (E) = \underbrace{S_0}_\text{vertical offset} + \underbrace{\frac{2 A_L}{\pi} \bigg[ \frac{\Delta E_L}{4(E - E_L)^2 + (\Delta E_L)^2}\bigg]}_\text{$S_L(E)$} + \underbrace{A_1 e^{-\frac{(E - E_1)^2}{(\Delta E_1)^2}} +  A_2 e^{-\frac{(E - E_2)^2}{(\Delta E_2)^2}}}_\text{$S_{PL}(E)$},
\end{equation}
where $E = \hbar \omega$ is the photon energy, $S_0$ is a vertical offset, $A_L$ is the amplitude of the laser peak, $E_L$ is the laser peak photon energy, $\Delta E_L$ is the laser peak width, $A_i$ are the amplitudes of the PL contributions, $E_i$ are the and PL peak photon energies, and $\Delta E_i$ are the PL peak widths, for $i = 1,2$. The empirical parameters used in fitting the PL spectral lineshapes to Eq. \ref{eqdetectorlineshapes} are listed in Tab. \ref{tab:spectraparameters}.

\begin{table}[htbp]
 \caption{\label{tab:spectraparameters}Empirical 2PE-PL spectra parameters in Eq. \ref{eqdetectorlineshapes}.}
  \begin{tabular}{@{}llll@{}}
    \hline
   Parameter & 1.8 nm nc-SiQDs & 2.3 nm nc-SiQDs & RhB \\
    \hline
   $A_1$ & 1.000 & 0.190 & 0.521\\
   $A_2$ & 0.620 & 1.000 & 1.000\\
   $E_1$ (eV) & 1.850 & 1.634 & 2.033\\
   $E_2$ (eV) & 2.021 & 1.770 & 2.119\\
   $\Delta E_1$ (eV) & 0.166 & 0.041 & 0.137\\
   $\Delta E_2$ (eV) & 0.146 & 0.173 & 0.053\\
    \hline
  \end{tabular}
\end{table}

\FloatBarrier

\subsection{\label{sec:dataarchiveerrorpropagation}Error Propagation of 2PA Cross Section}
The error in the 2PA cross section is propagated according to
\begin{equation}
\label{eqerrorpropagation}
|\delta \sigma_{samp}| = |\sigma_{samp}| \Bigg( \frac{|\delta(\sigma_{samp} n_0^{samp} z)|}{|(\sigma_{samp} n_0^{samp} z)|} + \frac{|\delta(\sigma_{ref} n_0^{ref} z)|}{|(\sigma_{ref} n_0^{ref} z)|} + \frac{|\delta \sigma_{ref}|}{|\sigma_{ref}|} + \frac{|\delta n_0^{ref}|}{| n_0^{ref}|} + \frac{|\delta n_0^{samp}|}{| n_0^{samp}|}\Bigg),
\end{equation}
where $samp$ refers to the nc-SiQD sample, $ref$ refers to the reference standard (RhB), $\sigma n_0 z$ is the corresponding fit parameter, $|\delta \sigma_{ref}| / |\sigma_{ref}| = 0.15$ in Makarov \textit{et al.},\cite{makarov} and $n_0$ is the corresponding measured number density.

\FloatBarrier
\subsection{\label{sec:dataarchivesummary}Summary of 2PA Parameters}

The values of the 2PA cross sections for the nc-SiQD samples and RhB reference standard are tabulated below. The maximum values of the figures of merit $\Delta_{ph}$ and $\Delta_m$ over a given intensity scan and the number of terms $N$ in Eq. 5 in the article required such that the 2PA cross section fit parameter varies by less than a threshold of 1\% are shown in Tab. \ref{tab:fom}. 

\begin{table}[ht]
\centering
 \caption{\label{tab:2pacrosssection}Summary of 2PA cross sections.}
  \begin{tabular}{c c | c c | c c | c c}
  \hline
   & & \multicolumn{2}{|c|}{nc-SiQD $d= 1.8 \pm 0.2$ nm} & \multicolumn{2}{|c|}{nc-SiQD $d= 2.3 \pm 0.3$ nm} & \multicolumn{2}{|c}{RhB\footnote{RhB reference standard data on 2PA cross section spectra used is from Makarov \textit{et al.}\cite{makarov} who report an uncertainty of $\pm 15\%$.}\cite{makarov}}\\ 
  $\lambda$ (nm)& $\hbar \omega$ (eV)& $\sigma$ (GM)& $\delta \sigma$ (GM)& $\sigma$ (GM)& $\delta \sigma$ (GM)& $\sigma$ (GM)& $\delta \sigma$ (GM)\\
  \hline
  650 & 1.91 & 310& 60& 2600& 600& 53& 8\\
  700 & 1.77 & 350& 70& 1700& 400& 240& 40\\ 
  750 & 1.65 & 170& 50& 900& 300& 67& 10\\ 
  800 & 1.55 & 66& 13& 290& 60& 120& 18\\ 
  850 & 1.46 & 70& 20& 260& 90& 180& 30\\ 
  1250 & 0.99 & 0.064& 0.017& 0.40& 0.10& 0.0052& 0.0008\\ 
  \hline
  \end{tabular}
\end{table}

\begin{table}[ht]
\centering
 \caption{\label{tab:fom}Summary of 2PA figures of merit $\mathrm{max} \Delta_{ph}$ and $\mathrm{max} \Delta_{m}$ for each intensity scan and the number of terms $N$ in Eq. 5 in the article necessary such that the variation in the fit parameter $\sigma$ is less than the threshold of 1\%.}
  \begin{tabular}{c c | c c c | c c c | c c c}
  \hline
   & & \multicolumn{3}{|c|}{nc-SiQD $d= 1.8 \pm 0.2$ nm} & \multicolumn{3}{|c|}{nc-SiQD $d= 2.3 \pm 0.3$ nm} & \multicolumn{3}{|c}{RhB}\\ 
  $\lambda$ (nm)& $\hbar \omega$ (eV)& $\mathrm{max} \Delta_{ph}$& $\mathrm{max} \Delta_m$& N& $\mathrm{max} \Delta_{ph}$& $\mathrm{max} \Delta_m$& N& $\mathrm{max} \Delta_{ph}$& $\mathrm{max} \Delta_m$& N\\
  \hline
  650 & 1.91 & 0.082& 0.0018& 2& 0.22& 0.013& 4& 0.00072& 0.00034& 1\\
  700 & 1.77 & 0.16& 0.0048& 3 & 0.24& 0.021& 4& 0.0057& 0.0037& 1\\ 
  750 & 1.65 & 0.12& 0.0057& 3& 0.21& 0.028& 4& 0.0025& 0.0025& 1\\ 
  800 & 1.55 & 0.034& 0.0012& 2& 0.053& 0.0052& 2& 0.0031& 0.0022& 1\\ 
  850 & 1.46 & 0.0067& 0.000052& 1& 0.0099& 0.00022& 1& 0.00088& 0.00014& 1\\ 
  1250 & 0.99 & 0.00029& 0.000073& 1& 0.00065& 0.00045& 1& 0.000022& 0.00022& 1\\ 
  \hline
  \end{tabular}
\end{table}

\FloatBarrier
\subsection{\label{sec:datarchiveplots}2PE-PL Intensity Scans}

The plots of photoluminescence vs. incident intensity are fit and the parameters used to determine the 2PA cross sections of the samples relative the reference standard, with known quantum yields and concentrations. The value of $N_{PL}$ is determined by integrating the PL spectra. Data points are indicated with solid circles, corresponding fits to the simple quadratic model with dashed curves, and corresponding fits including higher-order terms in Eq. 5 in the article with solid curves. Samples are distinguished by color; nc-SiQDs $d = 1.8 \pm 0.2$ nm annealed at $1000^{\circ}$C (red), nc-SiQDs $d = 2.3 \pm 0.3$ nm annealed at $1100^{\circ}$C (blue), and RhB reference standard (green).

\begin{figure}
  \includegraphics[scale=0.75]{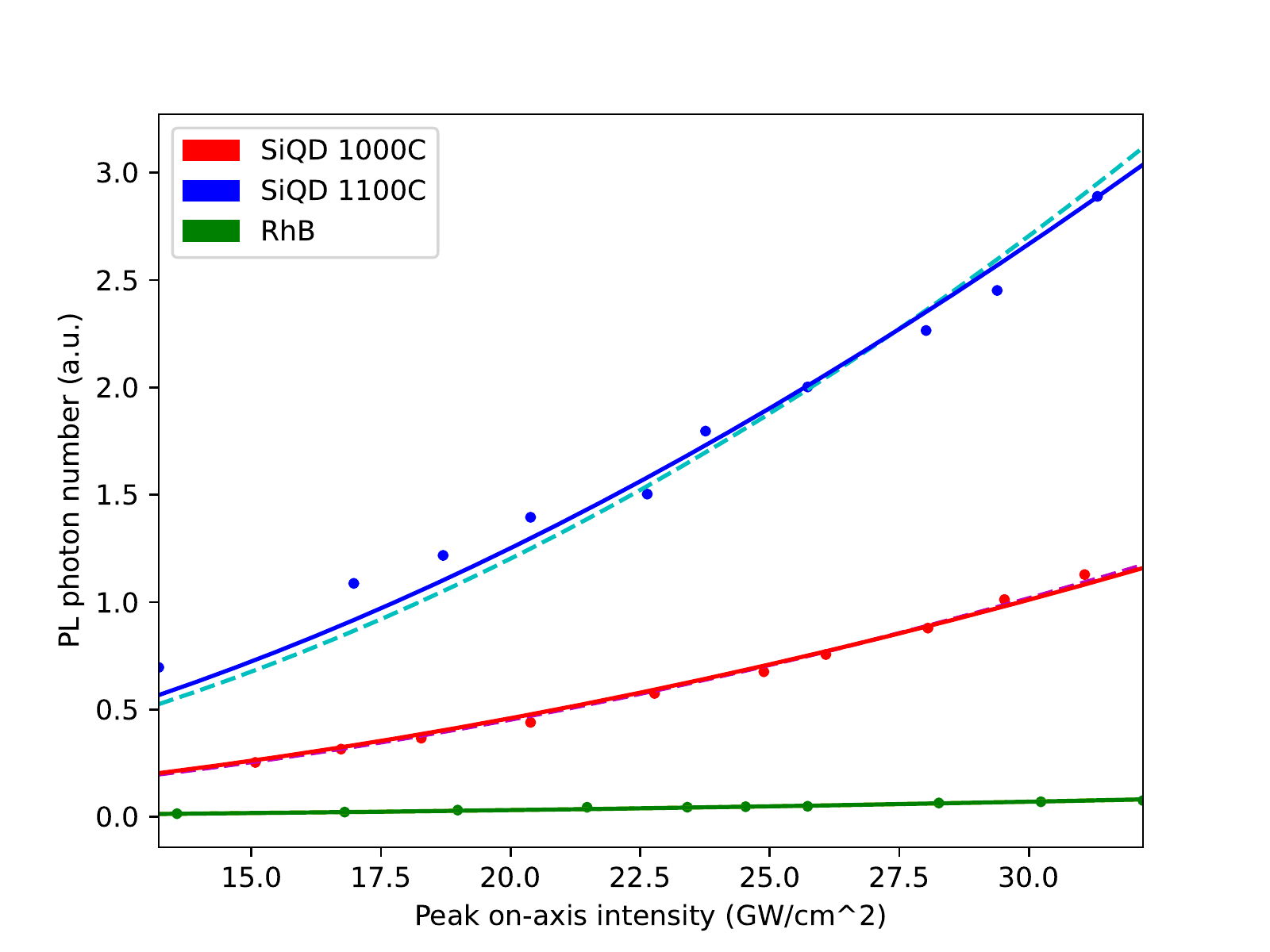}
  \caption{$N_{PL}$ vs. $I_0 (1-R)$ for all three samples at $\lambda = 650$ nm ($\hbar \omega = 1.91$ eV).}
  \label{fig:pli0650}
\end{figure}

\begin{figure}
  \includegraphics[scale=0.75]{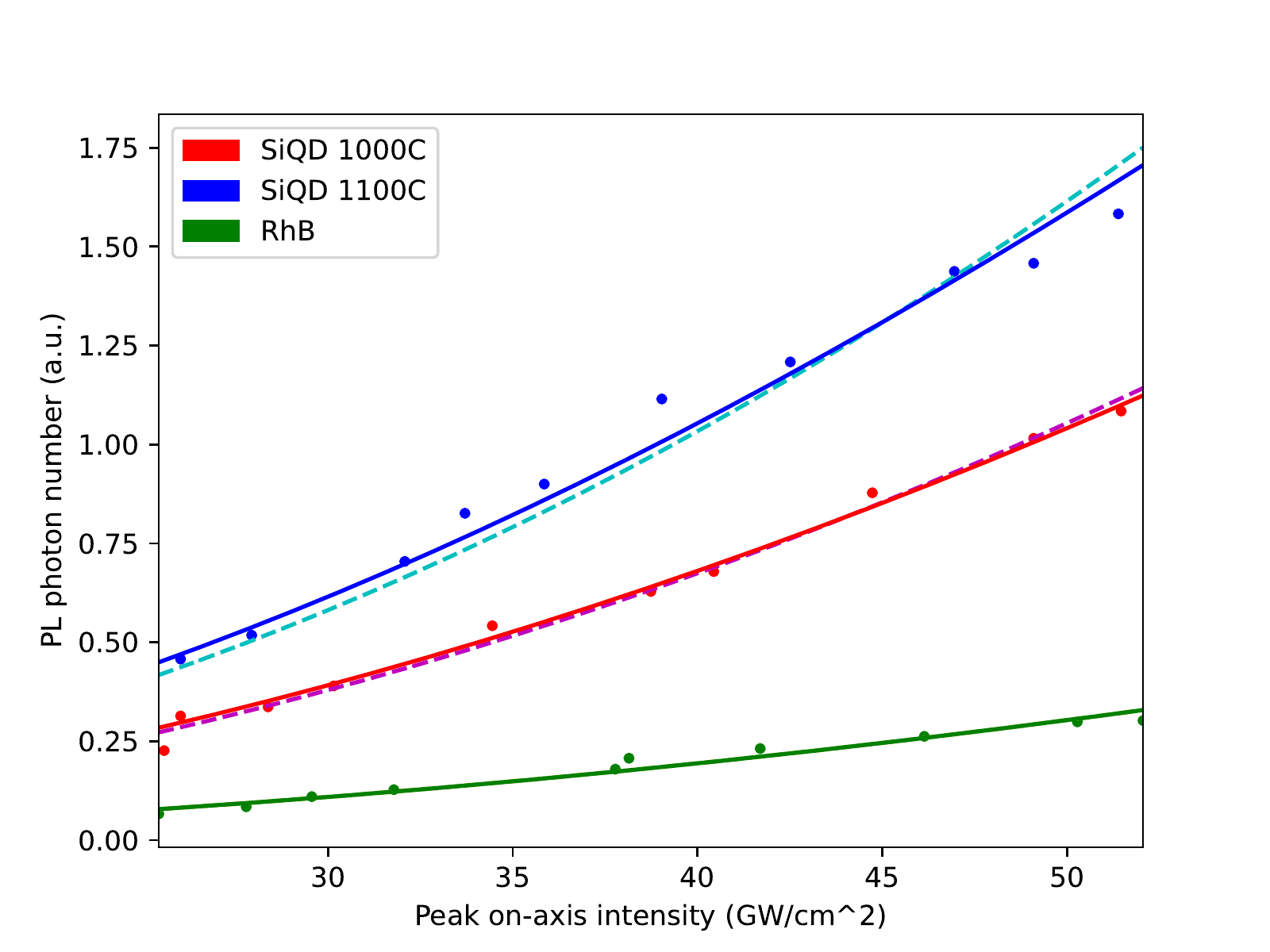}
  \caption{$N_{PL}$ vs. $I_0 (1-R)$ for all three samples at $\lambda = 700$ nm ($\hbar \omega = 1.77$ eV).}
  \label{fig:pli0700}
\end{figure}

\begin{figure}
  \includegraphics[scale=0.75]{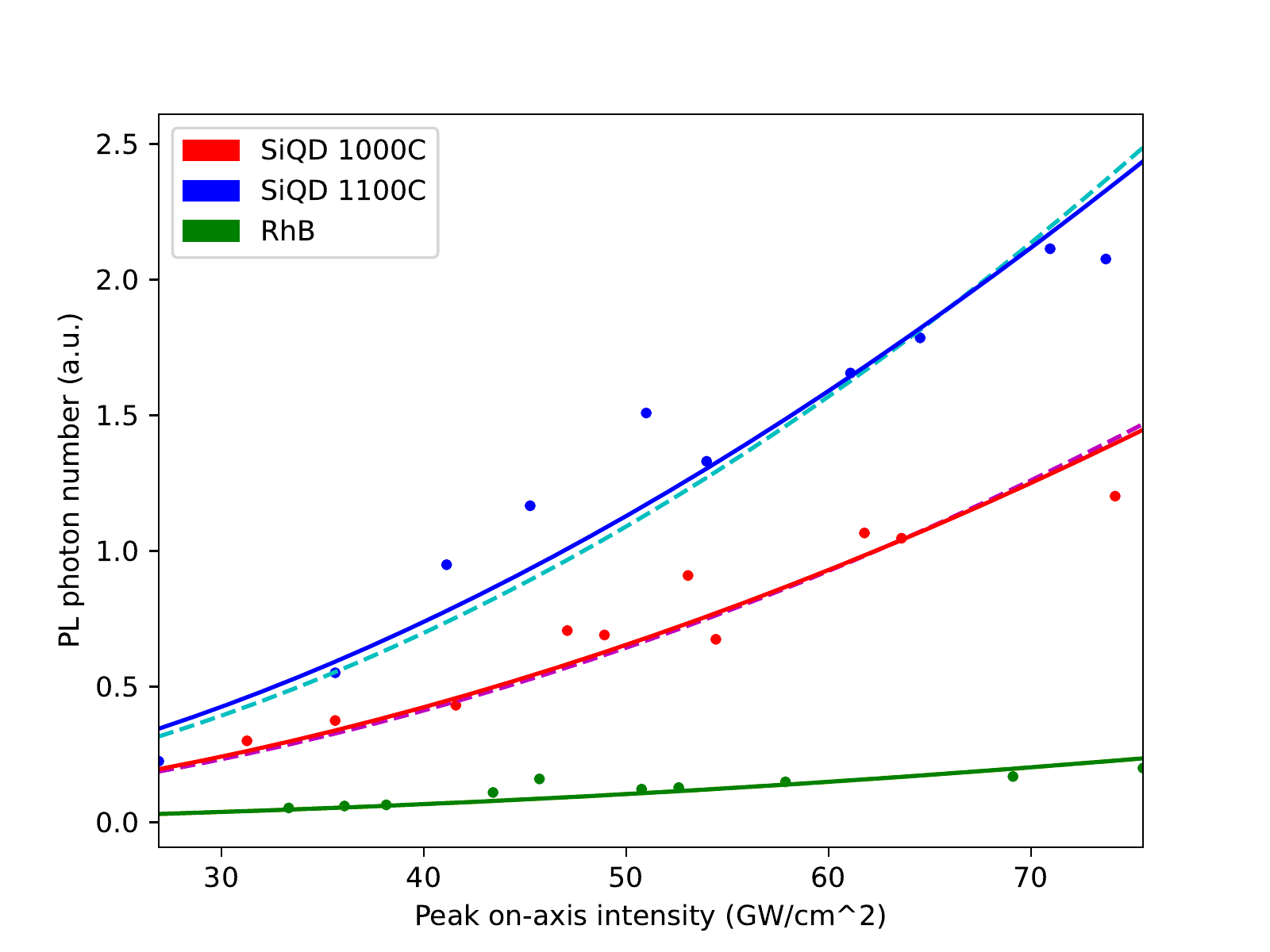}
  \caption{$N_{PL}$ vs. $I_0 (1-R)$ for all three samples at $\lambda = 750$ nm ($\hbar \omega = 1.65$ eV).}
  \label{fig:pli0750}
\end{figure}

\begin{figure}
  \includegraphics[scale=0.75]{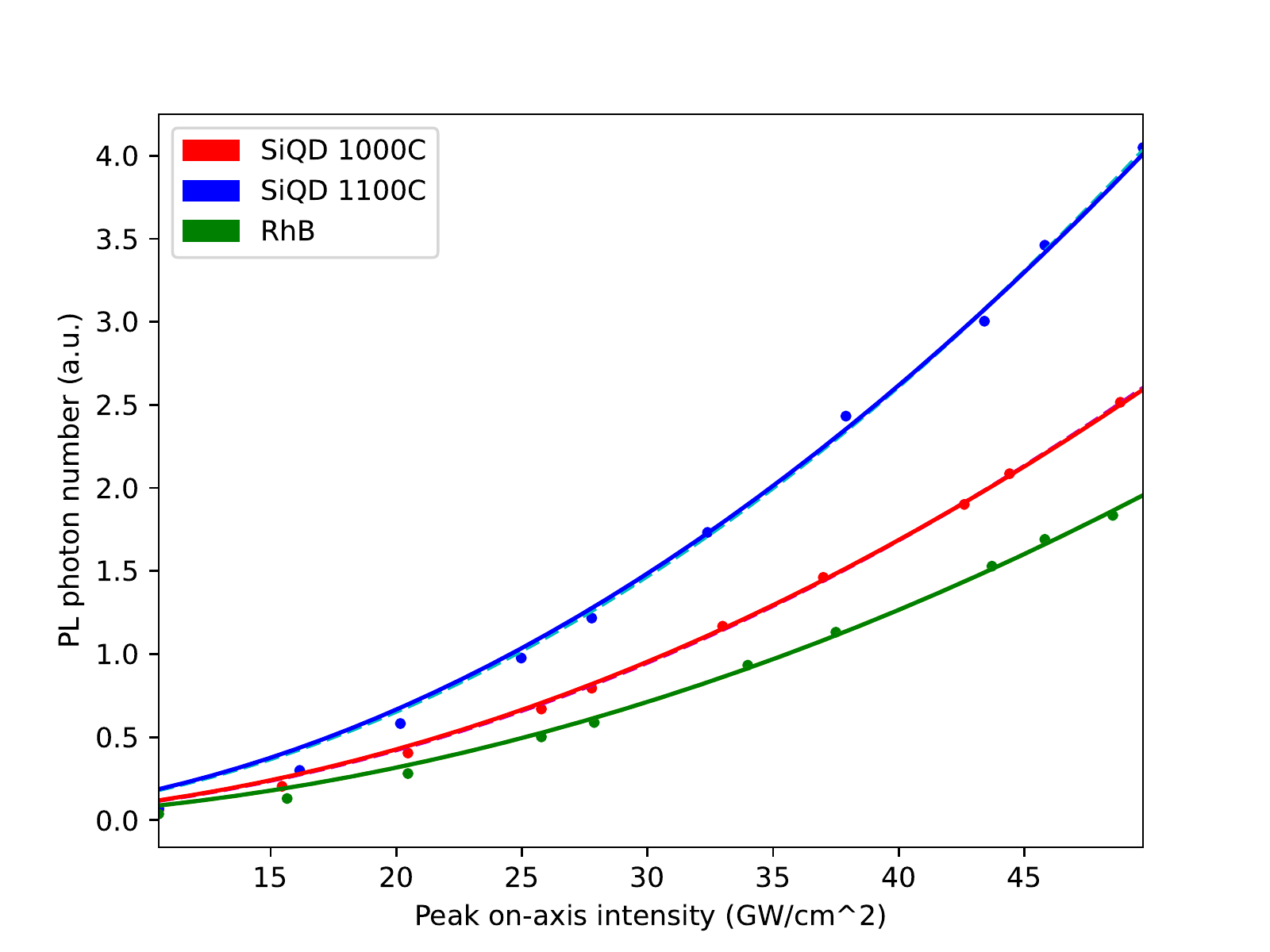}
  \caption{$N_{PL}$ vs. $I_0 (1-R)$ for all three samples at $\lambda = 800$ nm ($\hbar \omega = 1.55$ eV).}
  \label{fig:pli0800}
\end{figure}

\begin{figure}
  \includegraphics[scale=0.75]{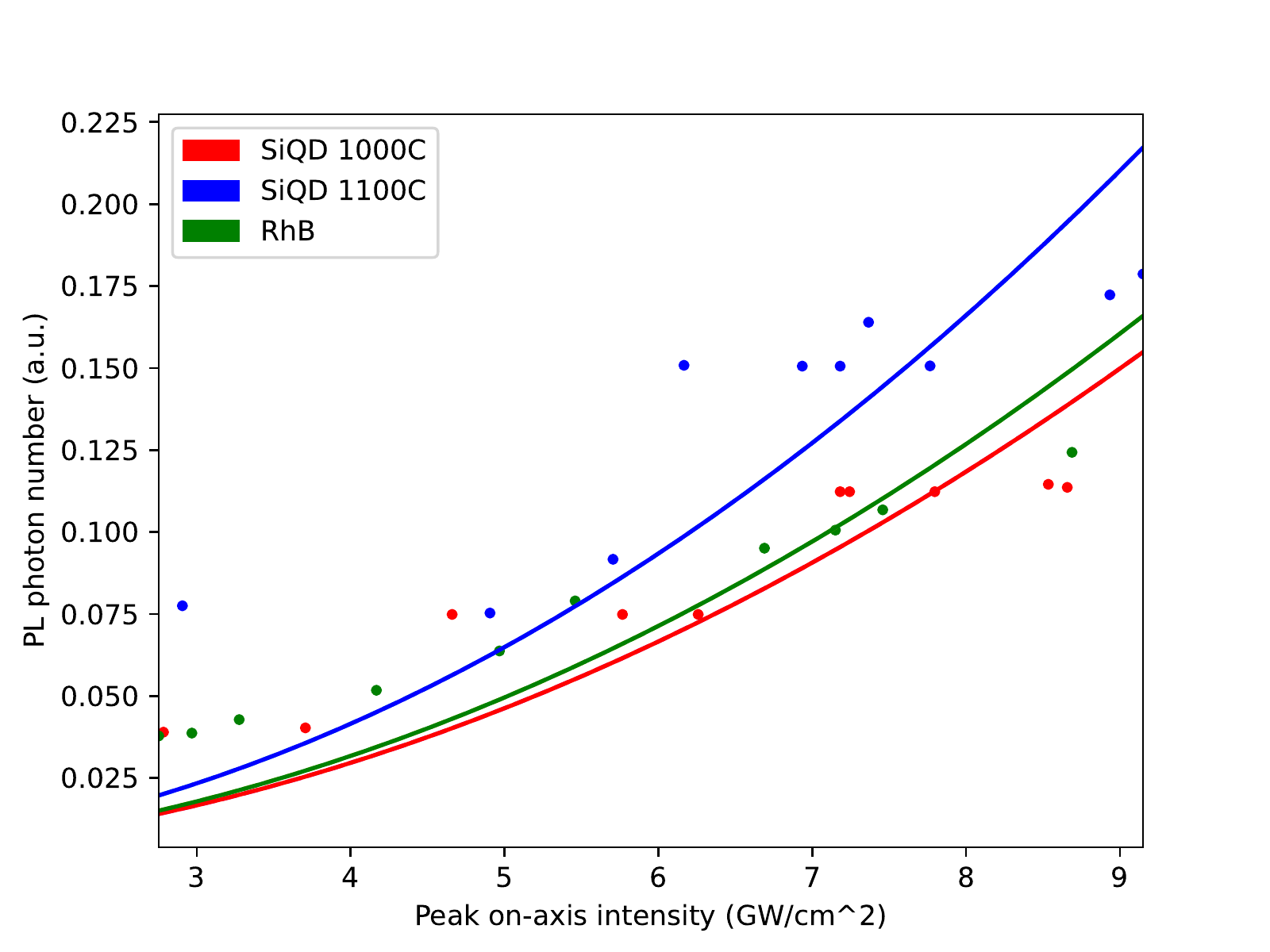}
  \caption{$N_{PL}$ vs. $I_0 (1-R)$ for all three samples at $\lambda = 850$ nm ($\hbar \omega = 1.46$ eV).}
  \label{fig:pli0850}
\end{figure}

\begin{figure}
  \includegraphics[scale=0.75]{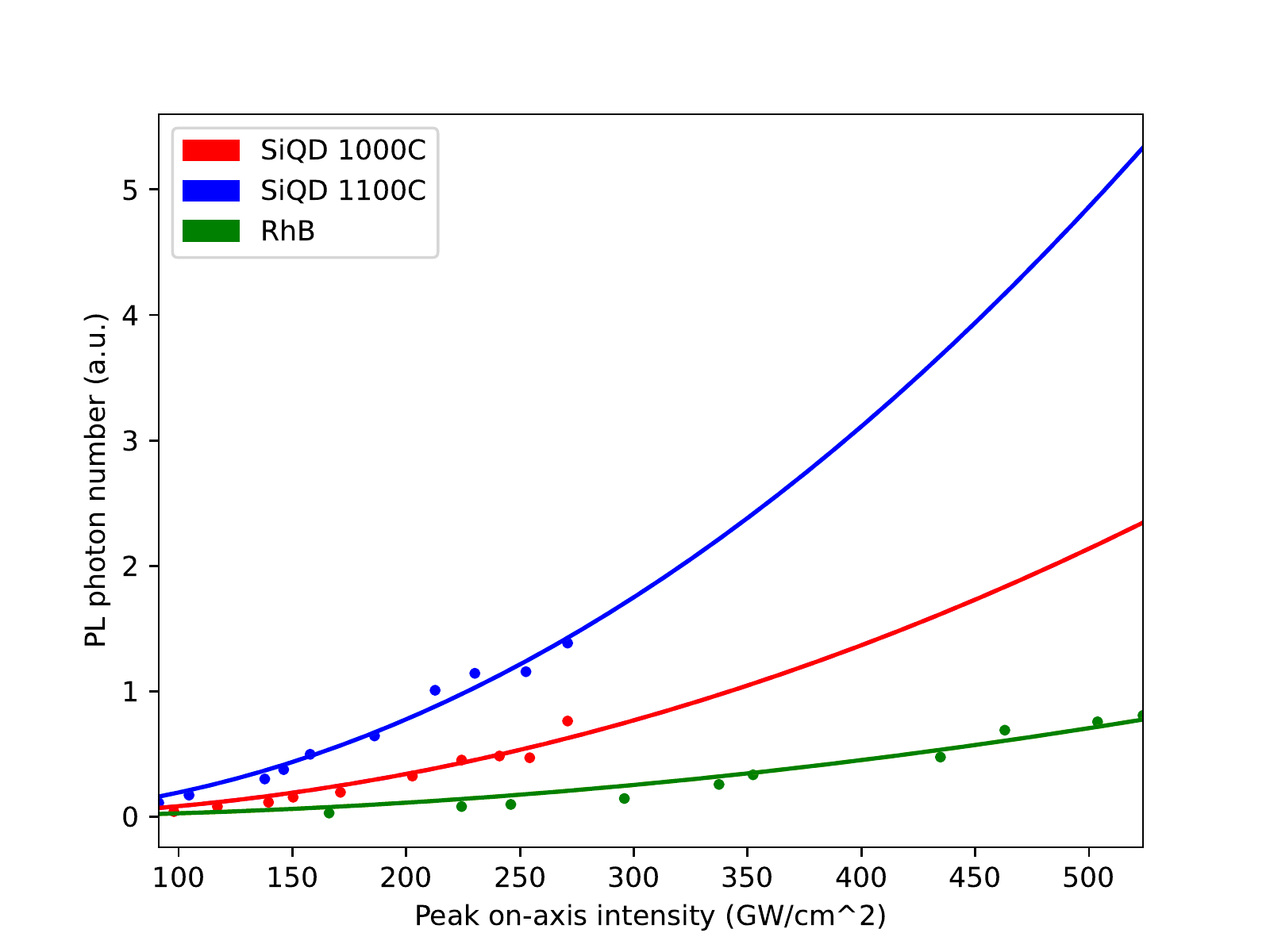}
  \caption{$N_{PL}$ vs. $I_0 (1-R)$ for all three samples at $\lambda = 1250$ nm ($\hbar \omega = 0.99$ eV).}
  \label{fig:pli01250}
\end{figure}

\FloatBarrier
\section{\label{sec:chi3molecular}Isotropic Molecular Third-Order Nonlinear Optical Susceptibility Tensor}

The 2PA cross section $\sigma(\omega)$ in the degenerate single-beam case is related to the imaginary part of the effective molecular third-order nonlinear optical susceptibility tensor $\mathrm{Im}\Big\{\xi^{(3)}_{XXXX}(\omega; \omega, \omega, -\omega)\Big\}$, also called the second hyperpolarizability, by\cite{kuzyk}
\begin{equation}
\label{eqsigmamoltensor}
\sigma(\omega) = \frac{3 \omega}{2 \varepsilon_0 \big[n(\omega)\big]^2 c^2} \mathrm{Im}\Big\{\xi^{(3)}_{XXXX}(\omega; \omega, \omega, -\omega)\Big\},
\end{equation}
where $\varepsilon_0$ is the vacuum permittivity, $n(\omega)$ the refractive index, and $c$ the speed of light. The 2PA response is isotropic due to the random orientation of nanocrystals in a colloidal suspension, and thus this tensor component is the rotational average of the molecular third-order nonlinear optical susceptibility tensor.\cite{craig} This relation is given by\cite{craig}
\begin{equation}
\label{eqmoltensor1}
\begin{split}
\xi^{(3)}_{XXXX}(\omega; \omega, \omega, -\omega) &= \Big\langle \boldsymbol\xi^{(3)}_{XXXX}(\omega;\omega, \omega,-\omega)\Big\rangle_{\phi,\theta,\psi}\\
&= \langle l^a_X l^b_X l^c_X l^d_X \rangle_{\phi,\theta,\psi}\ \boldsymbol\xi^{(3)}_{abcd}(\omega;\omega, \omega,-\omega)
\end{split}
\end{equation}
with $\boldsymbol\xi^{(3)}_{XXXX}(\omega; \omega, \omega, -\omega)$ fixed in space with the $X$-axis oriented along the incident electric field polarization, the lowercase latin indices corresponding to the axes fixed in the molecular frame, $\langle \cdots \rangle_{\phi,\theta,\psi}$ denoting the three-dimensional rotational average, and $l_A^a$ is the cosine of the angle between the space-fixed axis $A$ and the molecule-fixed axis $a$.

The rotational average of the direction cosines is rotationally invariant and can be expressed as a linear combination of isotropic rank-4 tensors. It is given by\cite{craig}
\begin{equation}
\label{eqrotationalaverageI}
\langle l^a_A l^b_B l^c_C l^d_D \rangle_{\phi,\theta,\psi} = \frac{1}{30}\begin{pmatrix}
\delta_{AB}\delta_{CD}\\
\delta_{AC}\delta_{BD}\\
\delta_{AD}\delta_{BC}\end{pmatrix}^T \begin{pmatrix}
4 & -1 & -1\\
-1 & 4 & -1\\
-1 & -1 & 4\end{pmatrix} \begin{pmatrix}
\delta_{ab}\delta_{cd}\\
\delta_{ac}\delta_{bd}\\
\delta_{ad}\delta_{bc}\end{pmatrix}.
\end{equation}
Crystalline Si has $m3m$ symmetry, for which there are four independent, non-vanishing tensor components; $\boldsymbol\xi_{aaaa}$, $\boldsymbol\xi_{aabb}$, $\boldsymbol\xi_{abab}$, and $\boldsymbol\xi_{abba}$ for $a \neq b$.\cite{popov} In the case of single-beam degenerate excitation, intrinsic permutation symmetry can be applied reducing this set to three independent components with $\boldsymbol\xi_{aabb} = \boldsymbol\xi_{abab}$.\cite{popov,boyd} Thus we have the result
\begin{equation}
\label{eqmoltensor2}
\xi^{(3)}_{XXXX}(\omega; \omega, \omega, -\omega) = \frac{1}{5} \Big[ 3 \boldsymbol\xi^{(3)}_{xxxx}(\omega; \omega, \omega, -\omega) + 4 \boldsymbol\xi^{(3)}_{xxyy}(\omega; \omega, \omega, -\omega) + 2 \boldsymbol\xi^{(3)}_{xyyx}(\omega; \omega, \omega, -\omega)\Big].
\end{equation}
As a result of the rotational averaging, it is the weighted average of tensor components given in Equation \ref{eqmoltensor2} that is directly related to the isotropic 2PA cross section in a single-beam 2PE-PL experiment of a colloidal suspension.

\FloatBarrier
\section{\label{sec:additionalsi}Additional Supporting Information}

Additional information that support the findings of this study are available from the corresponding author upon reasonable request.

\bibliography{supplement}% Produces the bibliography via BibTeX.